\documentclass[twocolumn,superscriptaddress,showpacs,preprintnumbers,amsmath,amssymb,aps,prb]{revtex4}
\usepackage{graphicx}
\usepackage{dcolumn}
\usepackage{bm}
\begin{document}
\title{Detrended fluctuation analysis of the magnetic and electric field variations that precede rupture}
\author{P. A. Varotsos}
\email{pvaro@otenet.gr}
\affiliation{Solid State Section and Solid Earth Physics Institute, Physics Department, University of Athens, Panepistimiopolis, Zografos 157 84,
Athens, Greece}
\author{N. V. Sarlis}
\affiliation{Solid State Section and Solid Earth Physics Institute, Physics Department, University of Athens, Panepistimiopolis, Zografos 157 84,
Athens, Greece}
\author{E. S. Skordas}
\affiliation{Solid State Section and Solid Earth Physics Institute, Physics Department, University of Athens, Panepistimiopolis, Zografos 157 84,
Athens, Greece}

\begin{abstract}
Magnetic field variations are detected before rupture in the form
of `spikes' of alternating sign. The distinction of these `spikes'
from random noise is of major practical importance, since it is
easier to conduct magnetic field measurements than electric field
ones.  Applying detrended fluctuation analysis (DFA), these
`spikes' look to be random at short time-lags. On the other hand,
long range correlations prevail at time-lags larger than the
average time interval between consecutive `spikes' with a scaling
exponent $\alpha$ around 0.9. In addition, DFA is applied to
recent preseismic electric field variations of long duration
(several hours to a couple of days) and reveals a scale invariant
feature with an exponent $\alpha \approx 1$ over all scales
available (around five orders of magnitude).

\end{abstract}

\maketitle

{\bf Many physical and biological complex systems exhibit
scale-invariant features characterized by long-range power-law
correlations, which are often difficult to quantify due to the
presence of erratic fluctuations, heterogeneity and
nonstationarity embedded in the emitted signals. Here, we focus on
different types of nonstationarities such as random spikes and
pseudo-sinusoidal trends, that may affect the long-range
correlation properties of signals that precede rupture. Since
these nonstationarities may either be epiphenomena of external
conditions or may arise from the intrinsic dynamics of the system,
it is crucial to distinguish their origin. This is attempted in
the present paper for both the magnetic and the electric field
variations that appear before rupture by employing the detrended
fluctuation analysis (DFA) as a scaling method to quantify
long-range temporal correlations. In particular, for the magnetic
field variations which have usually the form of `spikes' of
alternating sign, we find that at short time scales they look to
be random (thus may then be confused with random noise), but at
larger scales long-range correlations prevail. As for the electric
field variations, that are usually superimposed on a
pseudo-sinusoidal background, upon using the longest time series
available to date (i.e., with duration up to a couple of days), a
scale-invariant feature over five orders of magnitude emerges with
an exponent close to unity.}

\section{Introduction}

The detrended fluctuation
analysis\cite{PEN93,p18,p19,TAQ95,tal00,Hu01,CHE02,CHE05,Hu05}
(DFA) is a novel method that has been developed to address the
problem of accurately quantifying long range correlations in
non-stationary fluctuating signals.  It has been already applied
to a multitude of cases including
DNA\cite{OSS94,MAN94,MAN96,CAR02}, human motor activity\cite{HU04}
and gait\cite{HAUS01,ASH02}, cardiac
dynamics\cite{IVA99,HAV99,STAN99,IVA98},
meteorology\cite{IVANOVA99, IVANOVA03}, climate temperature
fluctuations\cite{KOS98}. Traditional methods such as  power
spectrum and autocorrelation analysis\cite{Stra81} are not
suitable for non-stationary signals\cite{tal00,Hu05}.

DFA is, in short, a modified root-mean-square (rms) analysis of a random walk and consists of the following steps:
 Starting with a signal $u(i)$, where $i=1,2,\ldots, N$, and $N$  is the length of the signal, the first step is to integrate $u(i)$ and obtain
 \begin{equation}
 y(i)=\sum_{j=1}^i \left[ u(j)-\bar{u} \right]
 \end{equation}
 where $\bar{u}$ stands for the mean
\begin{equation}
  \bar{u}=\frac{1}{N}\sum_{j=1}^N u(j).
\end{equation}
We then divide the profile $y(i)$ into boxes of equal length $n$.
In each box, we fit $y(i)$ using a polynomial function $y_n(i)$
which represents the local trend in that box. (If a different
order $l$ of polynomial fit is used, we have a different order
DFA-$l$, for example DFA-1 if $l$=1, DFA-2 if $l$=2, etc.) Next,
the profile $y(i)$ is detrended by substracting the local trend
$y_n(i)$ in each box of length $n$:
\begin{equation}
Y_n(i)=y(i)-y_n(i).
\end{equation}
Finally, the rms fluctuation for the integrated and detrended signal is calculated
\begin{equation}
F(n)\equiv \sqrt{\frac{1}{N} \sum_{i=1}^N \left[ Y_n(i) \right]^2}.
\end{equation}
The behavior of $F(n)$ over a broad number of scales is obtained
by repeating the aforementioned calculation of $F(n)$ for varied
box length $n$. For scale invariant signals, we find:
\begin{equation}
F(n) \propto n^\alpha
\end{equation}
where $\alpha$ is the scaling exponent. If $\alpha=0.5$, the
signal is uncorrelated (white noise), while if $\alpha > 0.5$ the
signal is correlated.

By employing the DFA method it was found\cite{NAT02,WER05} that
long-range correlations exist in the original time-series of the
so called seismic electric signals (SES)  activities. These are
low frequency ($\lesssim 1$Hz) electric signals preceding
earthquakes\cite{proto,FRA90,uye,uye2,VAR05C,SAR08}, the
generation of which could be understood in the following context.
A change of pressure affects the thermodynamic parameters for the
formation, migration or activation, in general, of defects in
solids\cite{VARALEX82}. In an
 ionic solid doped with aliovalent
impurities a number of extrinsic defects is produced,\cite{KOS75,VAR81} due to
charge compensation, a portion of
which is attracted by nearby aliovalent impurities, thus
forming electric dipoles that can change their orientation in
space through a defect motion\cite{VARALEX80,VARALEX81}. Hence, pressure variations may
affect the thermodynamic parameters of this motion, thus resulting in a decrease
or increase\cite{VARALEX80p} of the relaxation time of these dipoles, i.e., their
(re)orientation is taking place faster or slower when an external
electric field is applied. When the pressure, or the stress in
general, reaches a {\em critical} value\cite{varbook} a
{\em cooperative} orientation of these electric dipoles occurs,
which results in the emission of a transient electric signal. This
may happen in the focal region of a (future) earthquake since it
is generally accepted that the stress gradually changes there {\em
before} rupture.

It has been shown\cite{NAT03,NAT03B} that SES activities are
better distinguished from electric signals emitted from man-made
sources, if DFA is applied to a signal after it has been analyzed
in a newly introduced time domain, termed natural time $\chi$. In
a time series comprising $N$ events, the natural time $\chi_k =
k/N$ serves as an index\cite{NAT02} for the occurrence of the
$k$-th event. The evolution of the pair ($\chi_k, Q_k$) is
studied\cite{NAT02,NAT03,NAT03B,NAT04,NAT05B,VAR05C,NAT06A}, where
$Q_k$ denotes a quantity proportional to the {\em energy} released
in the $k$-th event.
 For dichotomous
signals, which is frequently the case of SES activities, the
quantity $Q_k$ stands for the duration of
the $k$-th pulse.
 The normalized power spectrum $\Pi(\omega )\equiv | \Phi (\omega ) |^2 $ was
introduced, where
\begin{equation}
\label{eq3} \Phi (\omega)=\sum_{k=1}^{N} p_k \exp \left( i \omega
\frac{k}{N} \right)
\end{equation}
and $p_k=Q_{k}/\sum_{n=1}^{N}Q_{n}$, $\omega =2 \pi \phi$; $\phi$
stands for the {\it natural frequency}. In natural time
analysis, the properties of
$\Pi(\omega)$ or $\Pi(\phi)$ are studied  for natural frequencies
$\phi$
 less than 0.5, since in
this range of $\phi$, $\Pi(\omega)$  or $\Pi(\phi)$ reduces
 to a {\em characteristic function} for the
probability distribution $p_k$  in the context of probability
theory.
 When the system enters the
{\em critical} stage, the following relation
holds\cite{NAT02,VAR05C}:
\begin{equation}
\Pi ( \omega ) = \frac{18}{5 \omega^2} -\frac{6 \cos \omega}{5
\omega^2} -\frac{12 \sin \omega}{5 \omega^3}. \label{fasma}
\end{equation}
For $\omega \rightarrow 0$, Eq.(\ref{fasma}) leads
to\cite{NAT02}
\[ \Pi (\omega )\approx 1-0.07
\omega^2\] which reflects\cite{VAR05C} that the variance of $\chi$
is given by
\[ \kappa_1=\langle \chi^2 \rangle -\langle \chi \rangle
^2=0.07,\] where $\langle f( \chi) \rangle = \sum_{k=1}^N p_k
f(\chi_k )$.
  The entropy $S$ in the natural time-domain is defined
as\cite{NAT03B} \[ S \equiv  \langle \chi \ln \chi \rangle -
\langle \chi \rangle \ln \langle \chi \rangle,\] which  depends on
the sequential order of events\cite{NAT04}. It
exhibits\cite{NAT05B} concavity, positivity,
Lesche\cite{LES82,LES04} stability, and for SES activities (critical dynamics) its value is
smaller\cite{NAT03B} than the value $S_u (=\ln 2
/2-1/4\approx 0.0966$) of a ``uniform'' (u) distribution (as
defined in Refs. \cite{NAT03,NAT03B,NAT04}, e.g.
when all $p_k$ are equal or $Q_k$ are positive independent and
identically distributed random variables of finite variance. In
this case, $\kappa_1$ and $S$ are designated $\kappa_u(=1/12)$ and
$S_u$, respectively.). Thus, $S < S_u$. The same holds for the
value of the entropy obtained\cite{NAT05B,NAT06A} upon considering
the time reversal ${\mathcal T}$, i.e., ${\mathcal T}
p_k=p_{N-k+1}$, which is labelled by $S_-$.
In summary, the SES activities, in contrast to the signals produced by man-made electrical sources,
when analyzed in natural time
exhibit {\em infinitely} ranged temporal correlations\cite{NAT03,NAT03B} and  obey
the conditions\cite{NAT06A}:
\begin{equation}\label{eq1}
    \kappa_1 = 0.07
\end{equation}
and
\begin{equation}\label{eq2}
    S, S_- < S_u.
\end{equation}

For major earthquakes, i.e., with magnitude Mw6.5 or larger, SES
activities are accompanied\cite{PRL03} by detectable variations of
the magnetic field {\bf B}\cite{sar02}. These variations, which
are usually measured by coil magnetometers (see below), have the
form of `spikes' of alternating sign. It is therefore of interest
to investigate whether these `spikes' exhibit long range temporal
correlations. This investigation, which is of major importance
since only magnetic field data are usually available in most
countries\cite{KAR02B,MOL92,FRA90} (since it is easier to conduct
magnetic field measurements than electric field ones),  is made
here in Section II.

In the up to date applications of DFA, long-range correlations
have been revealed in SES activities of duration up to a few
hours\cite{NAT02,NAT03,NAT03B,WER05}. During the last few years,
however, experimental results related to some SES activities of
appreciably longer duration, i.e., from several hours to a couple
of days, have been collected. These data now enable the
investigation of scaling in a wider range of scales than hitherto
known. This provides an additional scope of the present paper and
is carried out in Section III. A discussion of the results
concerning the magnetic and electric data follows in Section IV.
Finally, our conclusions are presented in Section V.

\section{Magnetic field variations preceding rupture}

The measurements have been carried out by three DANSK coil
magnetometers (DMM) oriented along the three axes: EW, NS and
vertical. Details on the calibration of these magnetometers can be
found in the Supplementary Information of Ref.\cite{PRL03}. In
particular, this calibration showed that for periods larger than
around half a second, the magnetometers measure\cite{PJA} the time
derivative $dB/dt$ of the magnetic field and their output is
`neutralized' approximately 200ms after the `arrival' of a
Heaviside unit step magnetic variation. It alternatively means
that a signal recorded by these magnetometers should correspond to
a magnetic variation that has `arrived' at the sensor less than
200ms before the recording. The data were collected by a Campbell
21X datalogger with sampling frequency $f_{exp}=1$sample/s.

\begin{figure}
\includegraphics{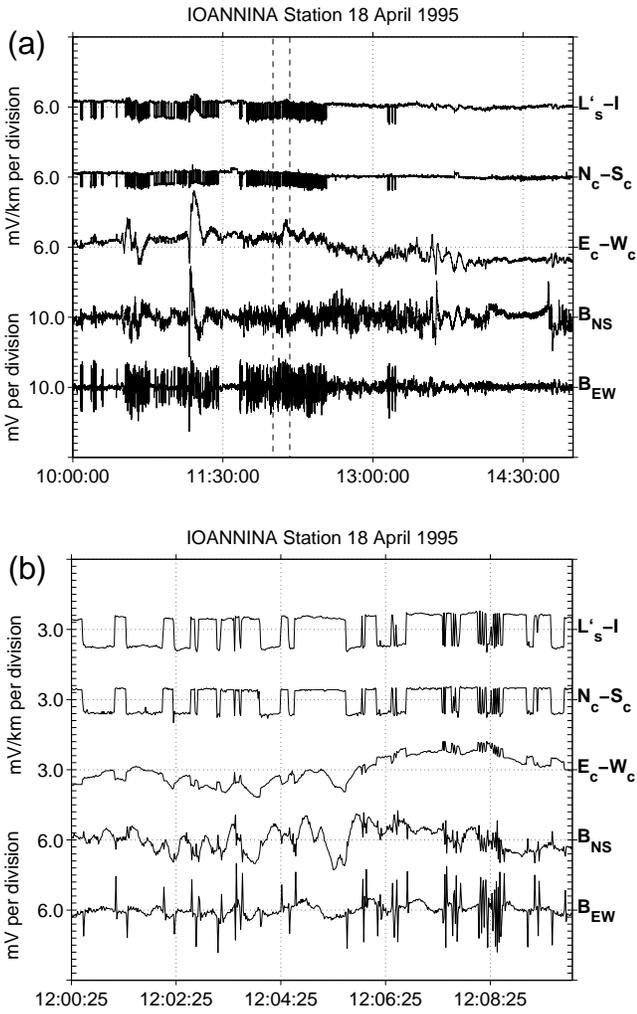}
\caption{(a):Variations of the  electric- (upper three channels)
and magnetic- (lower two channels) field recorded on April 18,
1995 (see the text). (b):A 10min excerpt of (a). The two dashed
lines in (a) show the excerpt depicted in (b).} \label{fig1m}
\end{figure}

Figure \ref{fig1m}(a) provides an example of simultaneous
recordings on April 18, 1995 at a station located close to
Ioannina (IOA) city in northwestern Greece. Variations of both the
electric (E) and magnetic field are shown. They were followed by a
magnitude Mw6.6 earthquake (according to the Centroid Moment
Tensor solutions reported by the United States Geological Survey)
with an epicenter at 40.2$^o$N21.7$^o$E that occurred almost three
weeks later, i.e., on May 13, 1995. The recordings of the two
horizontal magnetometers oriented along the EW- and NS-directions
labelled $B_{EW}$ and $B_{NS}$, respectively, are shown in the
lower two channels. They consist of a series of `spikes' of
alternating sign as it becomes evident in a 10min excerpt of these
recordings depicted in Fig.\ref{fig1m}(b). These `spikes' are
superimposed on a background which exhibits almost
pseudo-sinusoidal variations of duration much larger than 1s that
are induced by frequent small variations of the Earth's magnetic
field termed magnetotelluric variations (MT). In addition, the
horizontal $E-$variations were monitored at the same station by
measuring the variation $\Delta V$ of the potential difference
between (pairs of) electrodes -measuring dipoles- that are
grounded at depths of $\approx 2$m. Several such dipoles were
deployed along the EW- and NS-directions with lengths (L) a few to
several tens of meters (short-dipoles) or a couple of km
(long-dipoles) (Thus the electric field is given by $E=\Delta V/L$
and is usually measured in mV/km). For example, the following
measuring dipoles are shown in the upper three channels of
Fig.\ref{fig1m}(a): Two short electric dipoles at site `c' of IOA
station (see  the Supplementary Information of Ref.\cite{PRL03} as
well as Ref.\cite{APL05} where the selection of site `c' has been
discussed) of length 50m (labelled E$_c$-W$_c$ and  N$_c$-S$_c$,
respectively) while the uppermost channel corresponds to a long
dipole (labelled L'$_s$-I) with length $\approx 5$km at an almost
NS-direction. As it becomes obvious in Fig.\ref{fig1m}(b), the $E$
variations consist of a series of almost rectangular pulses (cf.
the initiation and cessation of each rectangular pulse correspond
to two `spikes' with opposite sign in the $B$ recordings).

\begin{figure}
\includegraphics{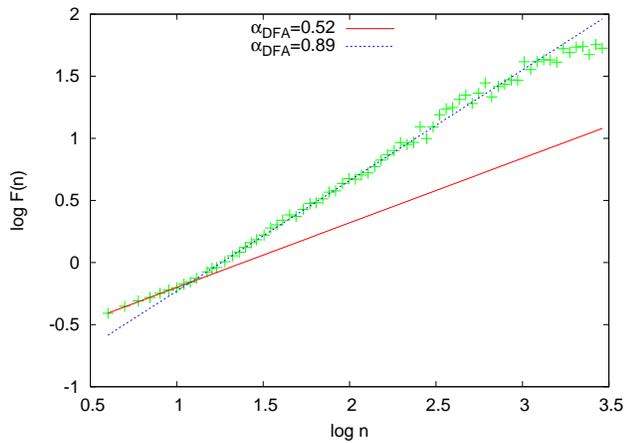}
\caption{(color online)The DFA for the $B_{EW}$ channel of
Fig.1(a). Logarithm base 10 is used.} \label{fig1mdfa}
\end{figure}

We now apply DFA to the original time-series of the magnetic field
variations and focus our attention on the $B_{EW}$ component where
the intensity of `spikes' is higher. Dividing the time series of
length $N$ into $N/n$ non-overlapping fragments, each of $n$
observations, and determining the local trend of the sub-series we
find the corresponding $\log F (n)$  versus $\log ( n )$ plot
where $n=f_{exp} \Delta t$ shown in Fig.\ref{fig1mdfa}. If we fit
the data with two straight lines (which are also depicted in
Fig.\ref{fig1mdfa}), the corresponding scaling exponents are
$\alpha=0.52\pm0.04$ and $\alpha=0.89\pm0.03$ for the short- and
long- time lags (i.e., smaller than around 12s and larger than
$\approx 12$s), respectively. The crossover occurs at a value
($\Delta t \approx 12$s) which is roughly equal to the average
duration  $\langle T \rangle \approx 11.01 \pm 0.03$s of each
electric pulse, corresponding to the interval between two
consecutive alternating `spikes'. Thus, Fig.\ref{fig1mdfa} shows
that at time-lags $\Delta t$ larger than $\langle T \rangle$
long-range power law correlations prevail (since $\alpha > 0.5$),
while at shorter time-lags the $\alpha$ value is very close to
that ($\alpha=0.5$) of an uncorrelated signal (white noise).

The above findings are strikingly reminiscent of the case of
signals with superposed {\em random} spikes studied by Chen et
al.\cite{CHE02}. They reported that for a correlated signal with
spikes, they found a crossover from uncorrelated behavior at small
scales to correlated behavior at large scales with an exponent
close to the exponent of the original stationary signal. They also
investigated the characteristics of the scaling of the spikes only
and found that the scaling behavior of the signal with {\em
random} spikes is a superposition of the scaling of the signal and
the scaling of the spikes. The case studied by Chen et
al.\cite{CHE02}, however,  is different from the present case,
because the `spikes' studied here  correspond to the pre-seismic
magnetic field variations and hence are {\em not} random (cf.
recall that when applying DFA to the `durations' of the electric
field rectangular pulses shown in Fig.\ref{fig1m}(b), we
found\cite{NAT03} an exponent around unity).

\section{DFA of SES activities of long duration}
\begin{figure}
\includegraphics{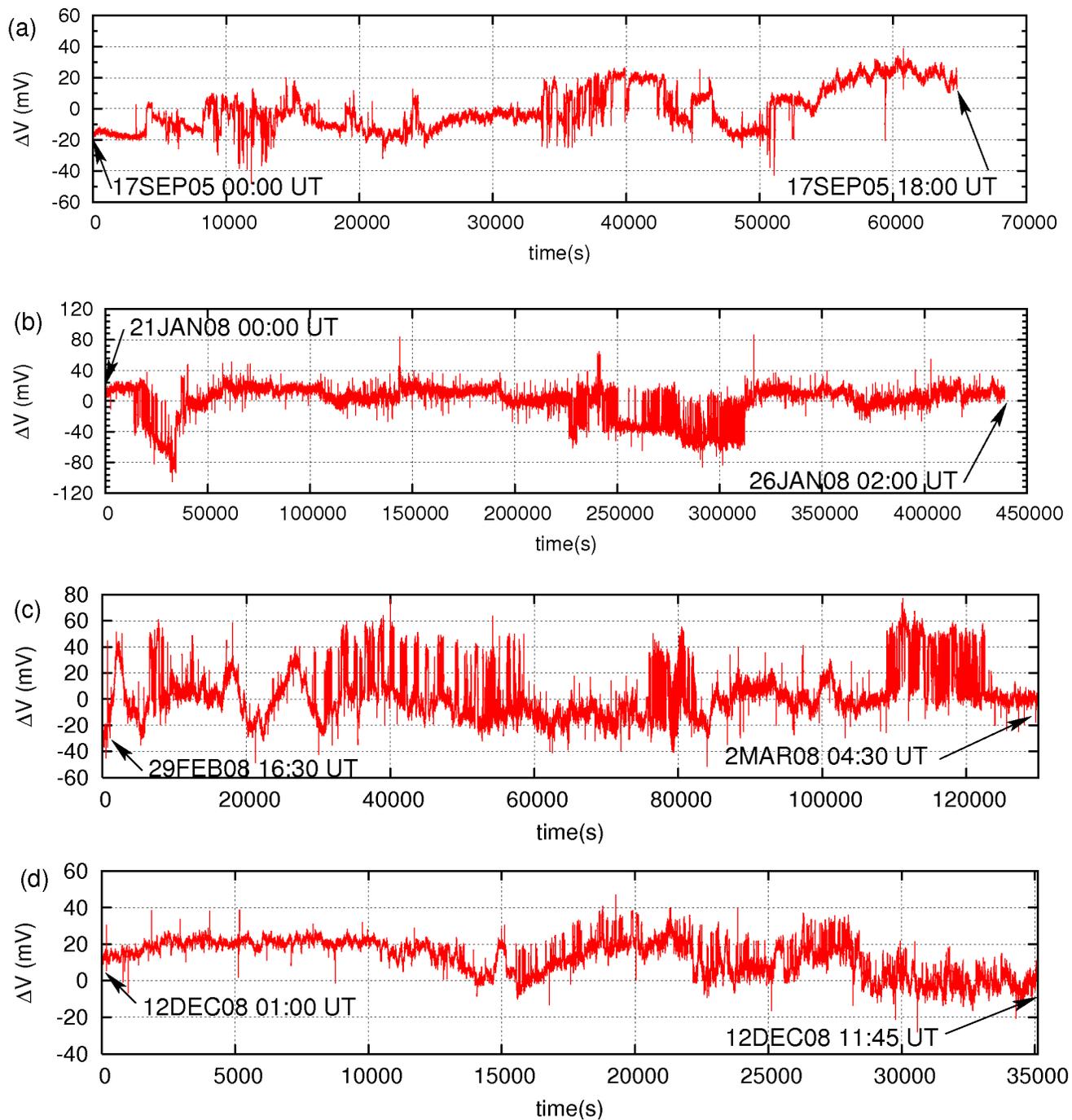}
\caption{(color online) Long duration SES activities during the
last few years (see the text).} \label{fig1}
\end{figure}

In Fig.\ref{fig1} the following four long duration SES activities
are depicted {\em all} of which have been recorded with sampling
frequency $f_{exp}=1$sample/s at a station close to Pirgos (PIR)
city located in western Greece (In this station only electric
field variations are continuously monitored with a multitude of
measuring dipoles the deployment of which has been described in
detail in the Supplementary Information of Ref.\cite{NAT06B}).
First, the SES activity on September 17, 2005 with duration of
several hours that preceded the Mw6.7 earthquake with an epicenter
at 36.3$^o$N23.3$^o$E on January 8, 2006. Second, the SES activity
that lasted from January 21 until January 26, 2008 and preceded
the Mw6.9  earthquake at  36.5$^o$N21.8$^o$E on February 14, 2008.
Third, the SES activity during the period from February 29 to
March 2, 2008 that was followed\cite{SAR08} by a Mw6.4 earthquake
at 38.0$^o$N21.5$^o$E on June 8, 2008 (SES activity information is
issued {\em only} when the expected magnitude is around 6 units or
larger\cite{NAT06A,NAT06B}). Finally, Fig.\ref{fig1}(d) depicts
the most recent SES activity of duration of several hours that was
recorded on December 12, 2008. The latter SES activity was
followed by a Mw5.4 earthquake at 37.1$^o$N20.8$^o$E on February
16, 2009 (according to Athens Observatory the magnitude is
Ms(ATH)=ML+0.5=6.0; we clarify that predictions are issued {\em
only} when the expected magnitude is comparable to 6.0 units or
larger). Note that this earthquake occurred after the initial
submission of the present paper (and its time of occurrence was
approached\cite{arxiv09a} by means of the procedure developed in
Ref.\cite{SAR08}).

\begin{figure}
\includegraphics{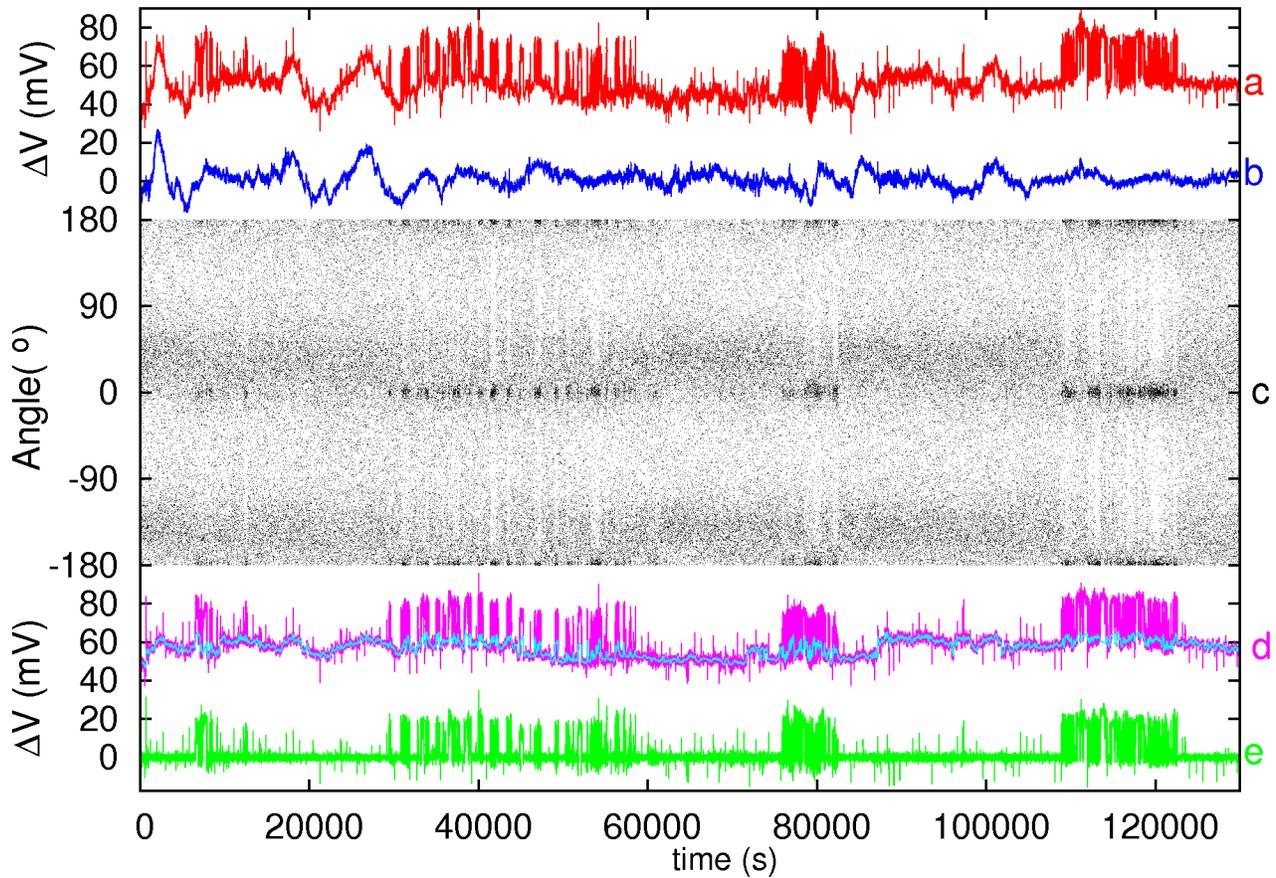}
\caption{(color online) The long duration SES activity from
February 29 to March 2, 2008. Channel ``a'':original time series,
channel ``b'': recordings at a measuring dipole which did not
record the SES activity, but does show MT variations, ``c'':the
angle of the resulting vector upon assuming that the ``1s
increments'' of channel  ``a'' lie along the x-axis and those of
channel ``b'' along the y-axis. Channel ``d'': the residual of a
linear least squares fit of channel ``a'' with respect to channel
``b'', channel ``e'':the same as ``d'' but after eliminating the
slight variations of the background. For the sake of clarity,
channels ``a'', ``b''  and ``d'' have been shifted vertically.}
\label{FigFeb}
\end{figure}

Here, we analyze, as an example, the long duration SES activity
that lasted from February 29 until March 2,
2008(Fig.\ref{fig1}(c)). The time-series of this electrical
disturbance, which is not of obvious dichotomous nature, is
depicted in the upper channel (labelled ``a'') of
Fig.\ref{FigFeb}. The signal, comprising a number of pulses, is
superimposed on a background which exhibits frequent small MT
variations. The latter are simultaneously recorded at {\em all}
measuring sites, in contrast to the SES activities that solely
appear at a restricted number of sites depending on the epicentral
region of the future earthquake\cite{varbook}. This difference
provides a key for their distinction. In order to separate the
background, we proceed into the following steps: First, we make
use of another measuring dipole of the same station, i.e., the
channel labelled ``b'' in Fig.\ref{FigFeb}, which has not recorded
the signal but does show the MT pseudo-sinusoidal variations.
Second, since the sampling rate of the measurements is 1sample/s,
we now compute the increments every 1s of the two time-series of
channels ``a'' and ``b''. Assuming the ``1s increments'' of
channel ``a'' lying along the x-axis and those of ``b'' along the
y-axis, we plot in  the middle panel  of Fig.\ref{FigFeb} (labeled
``c'') the angle of the resulting vector with dots. When a non-MT
variation (e.g., a dichotomous pulse) appears (disappears) in
channel ``a'', this angle in ``c'' abruptly changes to 0$^o$ ($\pm
180^o$). Thus, the dots in panel ``c'' mark such changes. In other
words, an increased density of dots (dark regions) around 0$^o$ or
$\pm 180^o$ marks the occurrence of these pulses, on which we
should focus. To this end, we plot in channel ``d'' of
Fig.\ref{FigFeb} the residual of a linear least squares fit of
channel ``a''  with respect to channel ``b''. Comparing channel
``d'' with channel ``a'', we notice a significant reduction of the
MT background but not of the signal. The small variations of the
MT background that still remain in ``d'', are now marked by the
light blue line, which when removed result in channel ``e'' of
Fig.\ref{FigFeb}. Hence, channel ``e'' solely contains the
electric field variations that precede rupture. This channel
provides the time-series which should now be analyzed.

\begin{figure}
\includegraphics{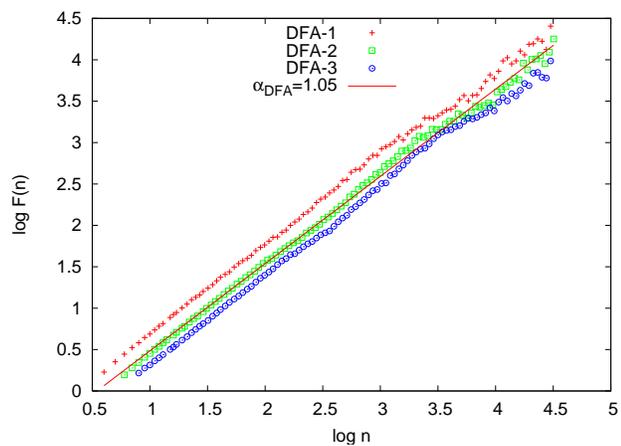}
\caption{(color online) The DFA-$l$ ($l$=1,2 and 3) for the lower
channel (i.e., the one labelled ``e'') of Fig.\ref{FigFeb}.
Logarithm base 10 is used.} \label{fig3}
\end{figure}

The DFA analysis (in conventional time) of the time-series of
channel ``e'' of Fig.\ref{FigFeb} is shown in Fig.\ref{fig3}. It
reveals an almost linear $\log F(n)$ vs $\log n$ plot (where
$n=f_{exp}\Delta t$) with an exponent $\alpha \approx 1$
practically {\em over all scales available} (approximately five
orders of magnitude). Note, that this value of the exponent
remains the same irrespective if we apply DFA-1, DFA-2 or DFA-3.
This result is in agreement with the results
obtained\cite{NAT02,NAT03,NAT03B,WER05} for SES activities of {\em
shorter} duration.

\begin{figure}
\includegraphics{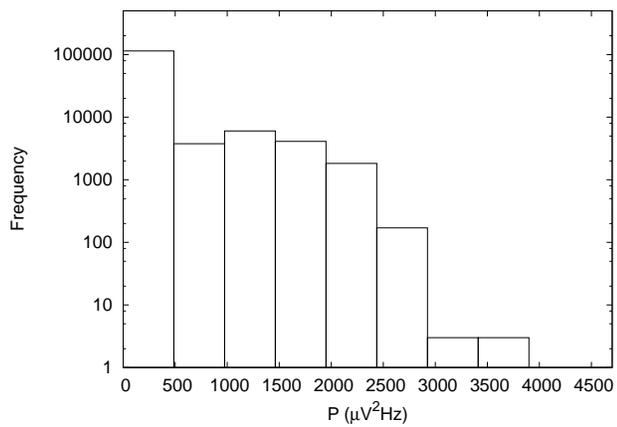}
\caption{Histogram of the ``instantaneous power'' P, i.e., the
squared amplitude of the signal depicted in channel ``e'' of
Fig.\ref{FigFeb}. } \label{FebHis}
\end{figure}

\begin{figure}
\includegraphics{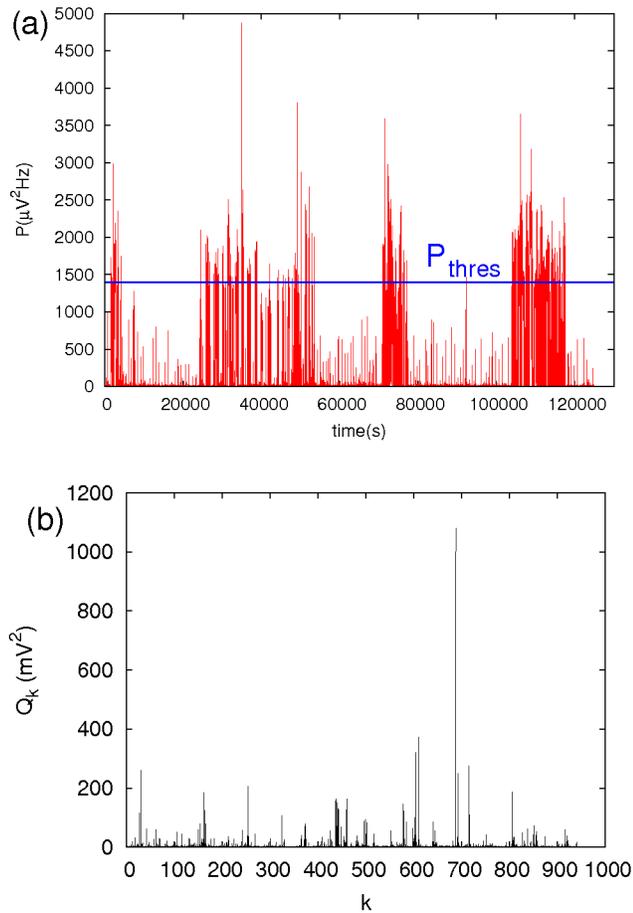}
\caption{(color online) (a) The``instantaneous power'' P of the
signal depicted in channel ``e'' of Fig.\ref{FigFeb} (computed by
squaring its amplitude). The solid line parallel to the x-axis
marks an example of a threshold P$_{thres}$(=1400 $\mu$V$^2$Hz)
chosen. (b)  The resulting representation of the signal depicted
in channel ``e'' of Fig.\ref{FigFeb} in natural time, when
considering P$_{thres}$=1400 $\mu$V$^2$Hz. } \label{FebExa}
\end{figure}

\begin{figure}
\includegraphics{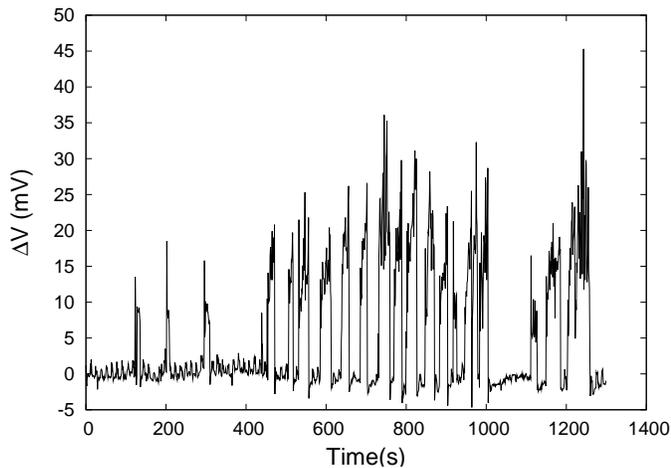}
\caption{The most recent electric signal analyzed (see the text).} \label{fig8}
\end{figure}

\begin{figure}
\includegraphics{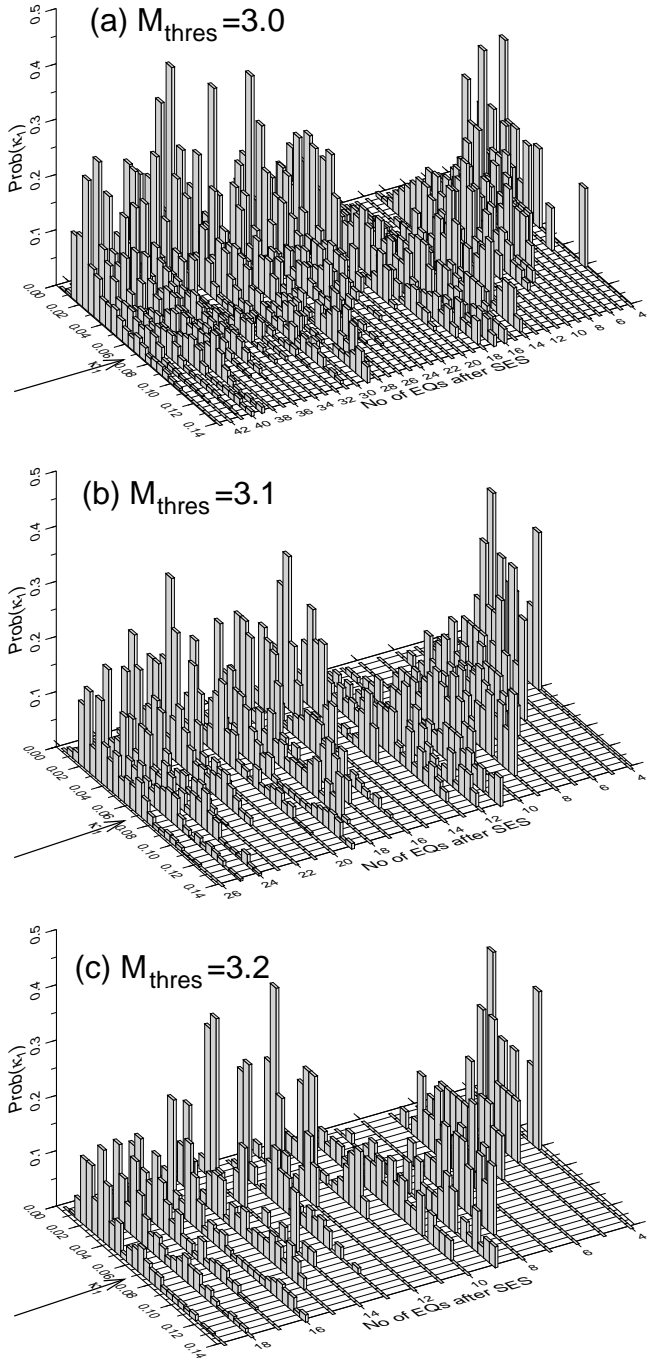}
\caption{The probability Prob($\kappa_1$) of the seismicity within
the area N$_{37.7}^{38.8}$E$_{22.6}^{24.1}$ subsequent to the SES
activity recorded at KER on March 28, 2009, when considering the
following magnitude thresholds: M$_{thres}$=3.0 (a), 3.1 (b) and
3.2 (c). } \label{fig9}
\end{figure}

\begin{figure}
\includegraphics{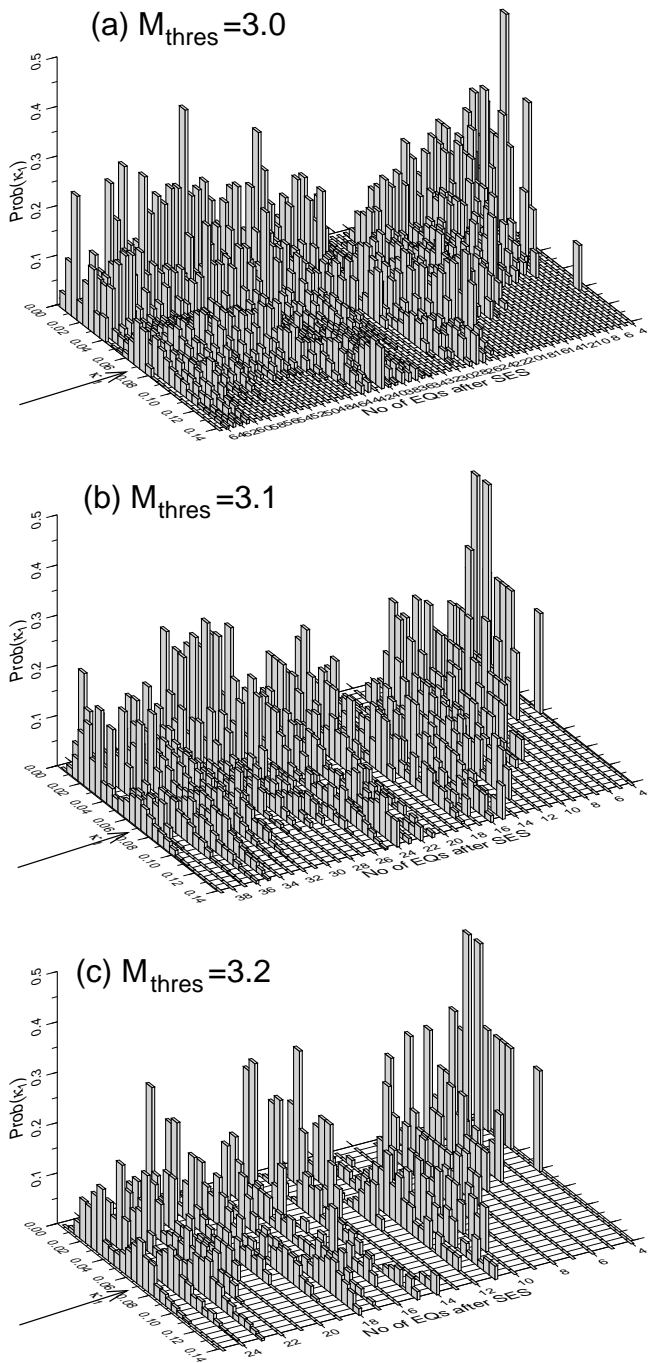}
\caption{The same as Fig.\ref{fig9}, but for the area
N$_{37.65}^{39.00}$E$_{22.20}^{24.10}$ in order to check the
spatial invariance of the results (see the text).} \label{fig10}
\end{figure}

For the classification of this signal, i.e., to distinguish
whether it is a true SES activity or a man-made electric signal,
we now proceed to its analysis in natural time. In order to define
($\chi_k, Q_k$) the individual pulses of the signal depicted in
channel ``e'' of Fig.\ref{FigFeb} have to be identified. A pulse
starts, of course, when the amplitude exceeds a given threshold
and ends when the amplitude falls below it. Moreover, since the
signal is not obviously dichotomous, instead of finding the
duration of each pulse, one should sum the ``instantaneous power''
during the pulse duration in order to find $Q_k$, as already
mentioned. To this end, we plot in Fig.\ref{FebHis} the histogram
of the instantaneous power P of channel ``e'' of Fig.\ref{FigFeb},
computed by squaring its amplitude. An inspection  of this figure
reveals a bimodal feature which signifies the periods of
inactivity  (P$<500\mu$V$^2$Hz) and activity (P$>500\mu$V$^2$Hz)
in channel ``e'' of Fig.\ref{FigFeb}. In order to find $Q_k$, we
focus on the periods of activity and select the power threshold
P$_{thres}$  around the second peak of the histogram in
Fig.\ref{FebHis}. Let us consider, for example, the case of
P$_{thres}$=1400 $\mu$V$^2$Hz. In Fig.\ref{FebExa}(a), we depict
the instantaneous power P of the signal in channel ``e'' of
Fig.\ref{FigFeb}. Starting from the beginning of the signal, we
compare P with P$_{thres}$ and when P exceeds P$_{thres}$ we start
summing up the P values until P falls below P$_{thres}$ for the
first time, $k=1$. The resulting sum corresponds to $Q_1$. This
procedure is repeated until P falls below P$_{thres}$ for the
second time, $k=2$, and the new sum represents $Q_2$, etc. This
leads to the natural time representation depicted in
Fig.\ref{FebExa}(b). The result depends, of course, on the proper
selection of P$_{thres}$. The latter can be verified by checking
whether a small change of P$_{thres}$ around the second peak of
the histogram, leads to a natural time representation resulting in
approximately the same values of the parameters $\kappa_1, S$ and
$S_-$. By randomly selecting P$_{thres}$ in the range 500 to 2000
$\mu$V$^2$Hz, we obtain  that the number of pulses in channel
``e'' of Fig.\ref{FigFeb} is $N=1100\pm 500$ with
$\kappa_1=0.070\pm 0.007$, $S=0.082\pm 0.012$ and $S_-=0.078\pm
0.006$. When P$_{thres}$ ranges between 1000 to 1500 $\mu$V$^2$Hz,
the corresponding values are $N=1200\pm 200$ with
$\kappa_1=0.068\pm 0.003$, $S=0.080\pm 0.005$ and $S_-=0.074\pm
0.003$. Thus, we observe that irrespective of the P$_{thres}$
value chosen, the parameters $\kappa_1, S$ and $S_-$ obey the
conditions (\ref{eq1}) and (\ref{eq2}) for the classification of
this signal as SES activity. The same holds for a non-dichotomous
 signal recorded on March 28, 2009 at Keratea station located
close to Athens (Fig.\ref{fig8}). To approach the occurrence time
of the impending event, the procedure developed in
Ref.\cite{SAR08} has been employed for the seismicity within the
area N$_{37.7}^{38.8}$E$_{22.6}^{24.1}$. The results, when
considering the seismicity until early in the morning on June 19,
2009, are shown in Fig.\ref{fig9}. They show that a local maximum
of the probability\cite{SAR08} Prob($\kappa_1$) of the seismicity
is observed at $\kappa_1=0.070$ upon the occurrence of the
following events: (a) an M$_L$=3.2 event at 12:42 UT on June 18
with epicenter at 38.7N 23.0E and an M$_L$=3.2 event at 03:44 UT
on June 19 at 37.9N 23.0E if we take a magnitude threshold
M$_{thres}$=3.0 (b) when taking M$_{thres}$=3.1, the maximum
occurs again upon the occurrence of the aforementioned event at
12:42 on June 18 (c) when increasing the threshold at
M$_{thres}$=3.2, the maximum occurs somewhat earlier, i.e., upon
the occurrence of the M$_L$=3.5 event at 08:26 UT on June 18 at
38.7N 23.0E. Let us now consider a larger area, i.e.,
N$_{37.65}^{39.00}$E$_{22.20}^{24.10}$, and repeat the same
calculation. The results are shown in Fig.\ref{fig10}. They reveal
again the occurrence of a local maximum of Prob($\kappa_1$) at
$\kappa_1=0.070$ upon the occurrence of the above mentioned
M$_L$=3.2 event at 12:42 UT on June 18 when considering either
M$_{thres}$=3.2 (Fig.\ref{fig10}(c)) or M$_{thres}$=3.1
(Fig.\ref{fig10}(b)) (for the latter case a maximum also occurs at
the last event shown in the figure with M$_L$=3.2 at 03:44 UT on
June 19). The maximum appears upon the occurrence of a M$_L$=3.0
at 23:48 UT on June 18 when decreasing the threshold at
M$_{thres}$=3.0.

In summary, the results exhibit both magnitude- and spatial-
invariance, which is characteristic when approaching the critical
point.

\begin{figure}
\includegraphics{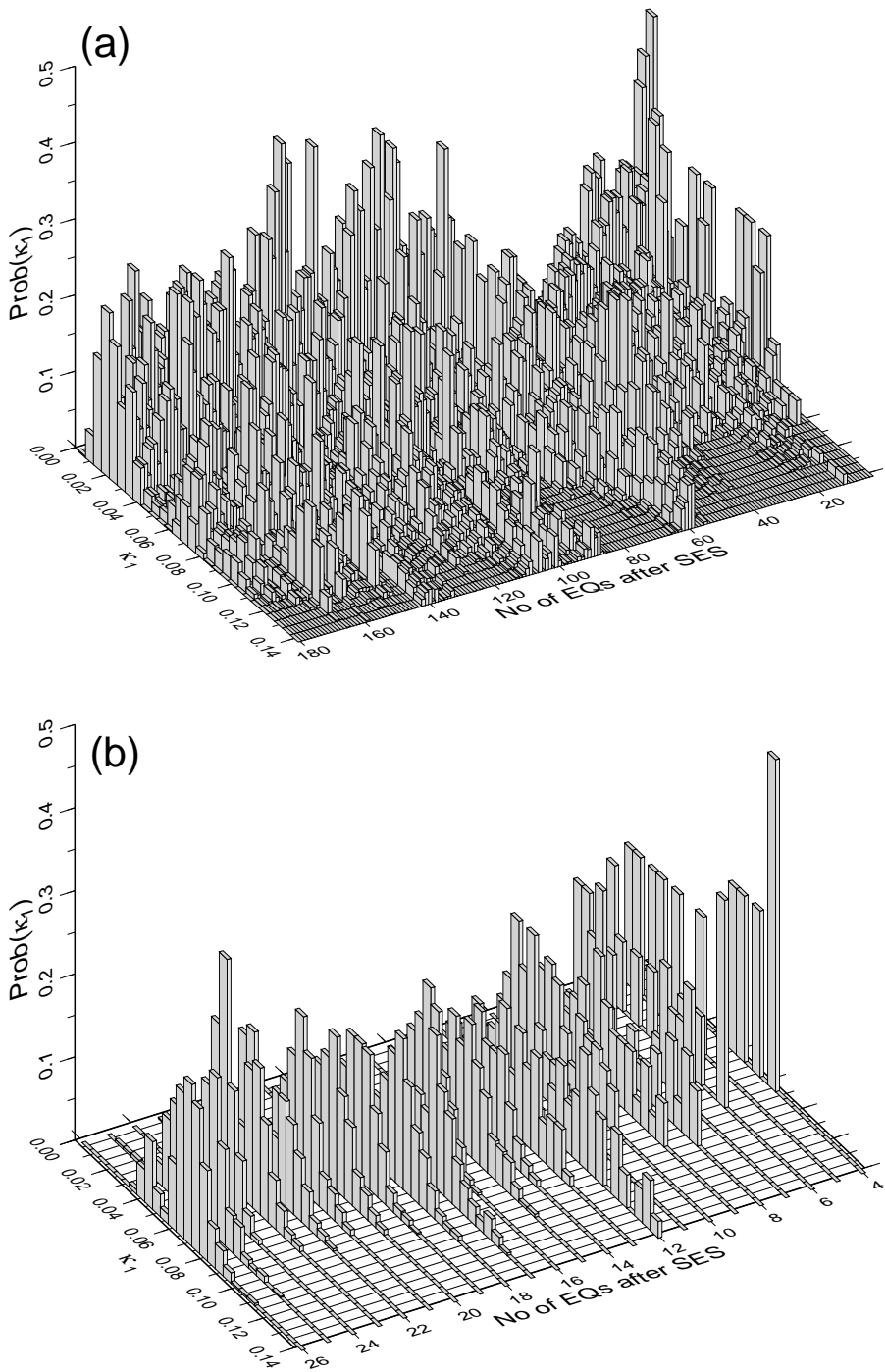}
\caption{The same as Fig.\ref{fig9}, but for the area
N$_{37.65}^{39.00}$E$_{22.20}^{24.10}$ by considering the
seismicity since (a) March 28, 2009 and (b) June 21, 2009 (see the
text).} \label{fig11}
\end{figure}

{\em Note added on July 21, 2009.} Repeating the aforementioned
calculation for the area N$_{37.65}^{39.00}$E$_{22.20}^{24.10}$,
we find the results depicted in Fig.\ref{fig11}(a) for
M$_{thres}$=2.8. An inspection of this figure reveals a local
maximum of Prob($\kappa_1$) at $\kappa_1=0.070$ upon the
occurrence of a M$_L$=3.0 at 21:10 UT on July 19, 2009 with an
epicenter at 38.1N 22.7E (the same result is found for
M$_{thres}$=2.9.) Interestingly, this local maximum turns to be
the prominent one when studying the seismicity in the same area
(for the same threshold M$_{thres}$=2.8) since June 21, 2009, see
Fig.\ref{fig11}(b); the same holds for M$_{thres}$=3.0.
Furthermore, the latter study was made for the smaller area
N$_{37.7}^{38.8}$E$_{22.6}^{24.1}$ for M$_{thres}$= 2.6, 2.7 and
3.0 and showed a local maximum of Prob($\kappa_1$) at
$\kappa_1=0.070$ upon the occurrence of the same event (i.e., at
21:10 UT on July 19, 2009). The investigation still continues in
order to clarify whether the latter local maximum will eventually
turn to become a prominent one.

\begin{figure}
\includegraphics{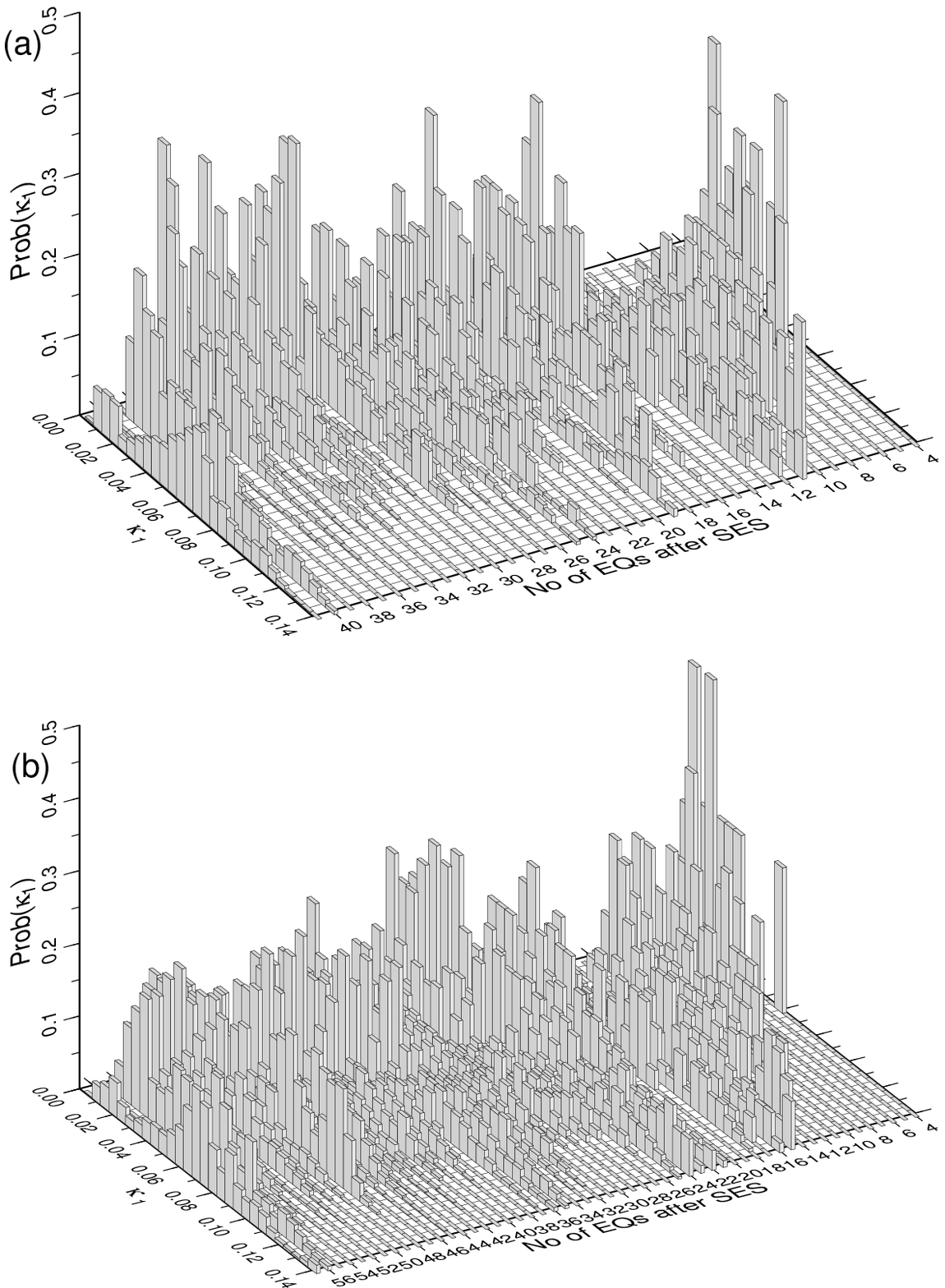}
\caption{The probability Prob($\kappa_1$) vs $\kappa_1$ upon
considering the seismicity (M$_{thres}$=3.1) until 9:35 UT on
September 2, 2009: Panel (a) corresponds to the area
N$_{37.7}^{38.8}$E$_{22.6}^{24.1}$, whereas panel (b) to the
larger area N$_{37.65}^{39.00}$E$_{22.20}^{24.10}$. }
\label{fig12}
\end{figure}

 {\em Note added on September 5, 2009.}  At 8:17 UT on September 2, 2009,
 a M$_L$=4.3 earthquake occurred almost at the center of the predicted area N$_{37.7}^{38.8}$E$_{22.6}^{24.1}$.
The continuation of the investigation of the previous paragraph in
the smaller area  N$_{37.7}^{38.8}$E$_{22.6}^{24.1}$ showed a
prominent maximum of   Prob($\kappa_1$) at $\kappa_1=0.070$ (see
Fig.\ref{fig12} (a)) upon the occurrence of a  M$_L$=3.3 event at
09:35 UT on September 2, 2009, with an epicenter at 38.1N 23.3E.
This was found to hold for the aforementioned larger area (see
Fig.\ref{fig12} (b)) as well, thus revealing that the {\em
criticality} condition -as developed in Ref. \onlinecite{SAR08}-
is obeyed.

\begin{figure}
\includegraphics{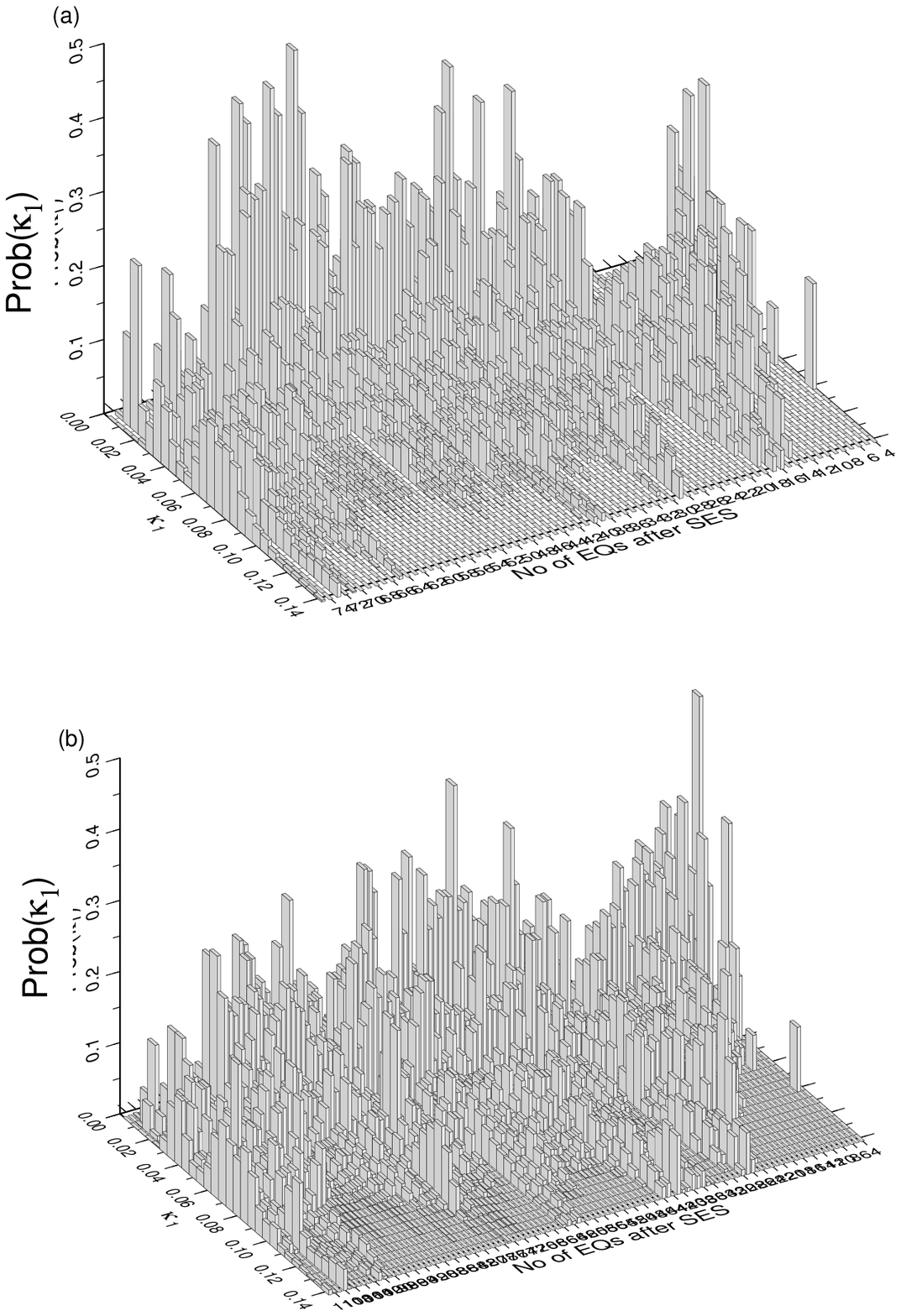}
\caption{The probability Prob($\kappa_1$) vs $\kappa_1$ upon
considering the seismicity (M$_{thres}$=3.0) until 1:55 UT on
September 18, 2009: Panel (a) corresponds to the area
N$_{37.7}^{38.8}$E$_{22.6}^{24.1}$, whereas panel (b) to the
larger area N$_{37.65}^{39.00}$E$_{22.20}^{24.10}$. }
\label{fig13}
\end{figure}

 {\em Note added on September 19, 2009.} The previous Note was followed by three earthquakes of magnitude 4.4, 4.7 and 4.8 that occurred
at 21:15 UT September 9, 06:04 UT September 10 and 07:12 UT
September 16, 2009 with epicenters at 38.7N 22.8E, 38.3N 24.1E and
39.0N 22.2E, respectively. The continuation of the natural time
investigation  of the seismicity for $M_{thres}$=3.0 in the
smaller area N$_{37.7}^{38.8}$E$_{22.6}^{24.1}$ revealed a
prominent maximum Prob($\kappa_1$) at $\kappa_1=0.070$ (see
Fig.\ref{fig13} (a)) upon the occurrence of a  M$_L$=3.0 event at
01:55 UT on September 18, 2009, with epicenter at 38.7N22.8E.
Since the same was found to be true for the aforementioned larger
area as well (see Fig.\ref{fig13} (b)), the investigation is now
extended to additional magnitude thresholds in order to clarify
whether (beyond spatial and) magnitude threshold invariance holds,
thus identifying that the system approaches the {\em critical}
point.

 {\em Note added on October 13, 2009.} The previous note was actually followed by an earthquake of magnitude 4.6 that occurred
at 22:27 UT on October 2, 2009 with an epicenter at 38.3N 22.9E.
The natural time investigation of the seismicity still continues
with the following results. First, in the smaller area
N$_{37.7}^{38.8}$E$_{22.6}^{24.1}$: For $M_{thres}$=2.7, the
criticality condition started to be approached since October 7 and
the {\em most prominent} maximum of Prob($\kappa_1$) at
$\kappa_1=0.070$ was observed upon the occurrence of the M$_L$=2.7
event at 01:56 UT on October 11, 2009. The same behavior was found
for $M_{thres}$=2.8 (but the last event for this case occurred at
08:15UT on October 10, 2009 at 38.2N 23.1E). Second, in the larger
area N$_{37.65}^{39.00}$E$_{22.20}^{24.10}$: For both thresholds
$M_{thres}$=2.7 and $M_{thres}$=2.8, the criticality condition was
also started to be approached since 20:18 UT on October 7, 2009.
The {\em most prominent} maximum of Prob($\kappa_1$) at
$\kappa_1=0.070$ for $M_{thres}$=2.7 have been observed upon the
occurrence of the aforementioned M$_L$=2.7 event at 01:56 UT on
October 11, 2009, while for  $M_{thres}$=2.8 it was observed at
4:51 UT on October 9, 2009, upon the occurrence of the M$_L$=2.9
event at 38.1E 23.2E. Interestingly, the results for the larger
area are almost identical if the investigation starts at 15:00 UT
on July 31, 2009.

\begin{figure}
\includegraphics{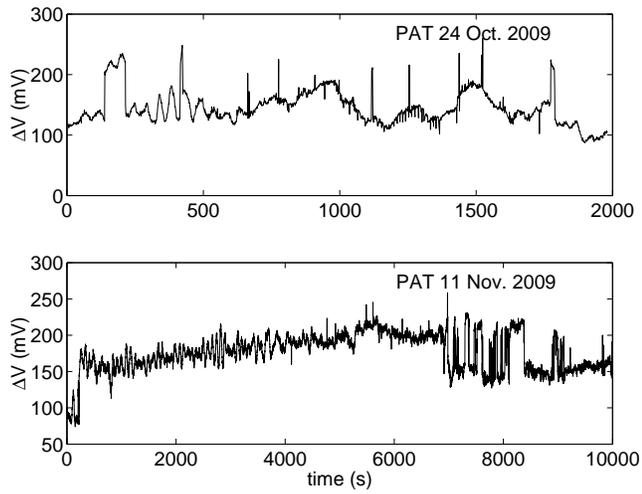}
\caption{The SES activities at PAT on 24 October 2009 (upper) and
11 November 2009 (lower), see the text.} \label{fig14}
\end{figure}

\begin{figure}
\includegraphics{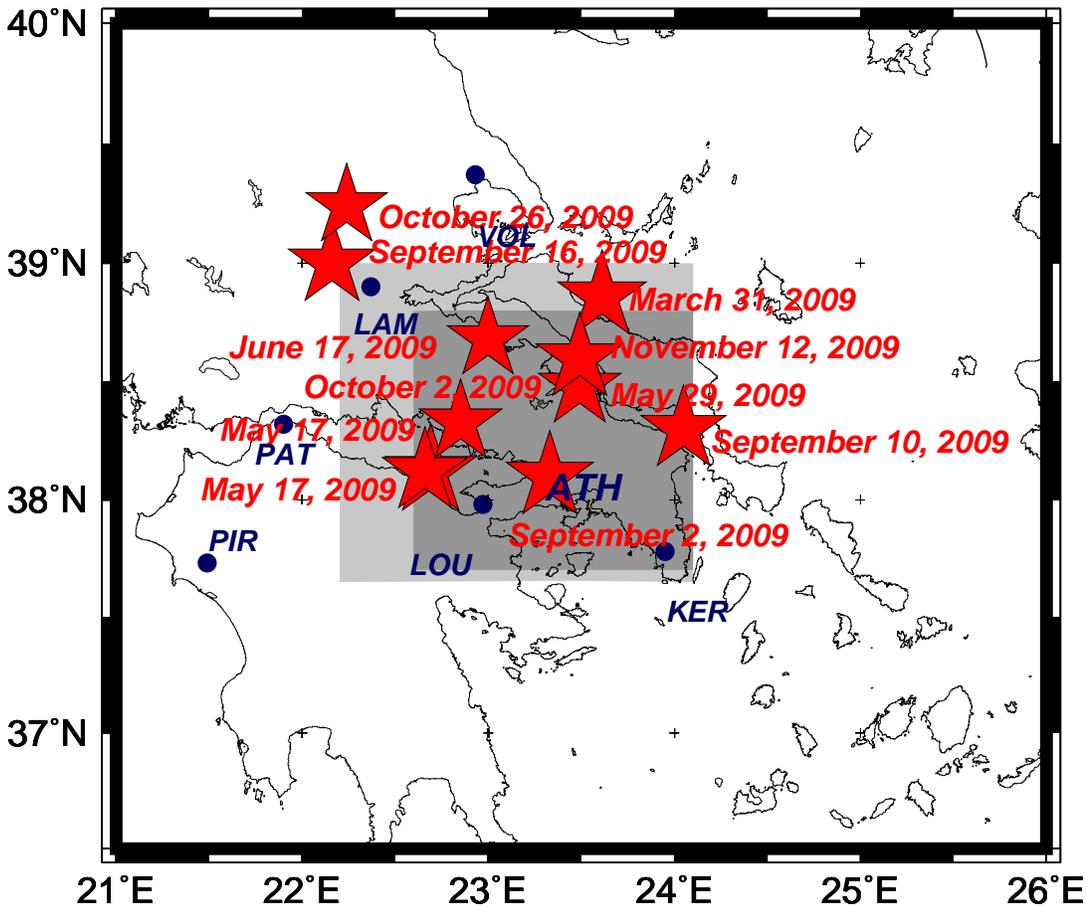}
\caption{Earthquakes with magnitude Ms(ATH)$\geq$4.5 and
epicenters within (or very close to) the area(s) specified in
advance that followed the consecutive Notes added to this paper.}
\label{fig15}
\end{figure}

\begin{figure}
\includegraphics{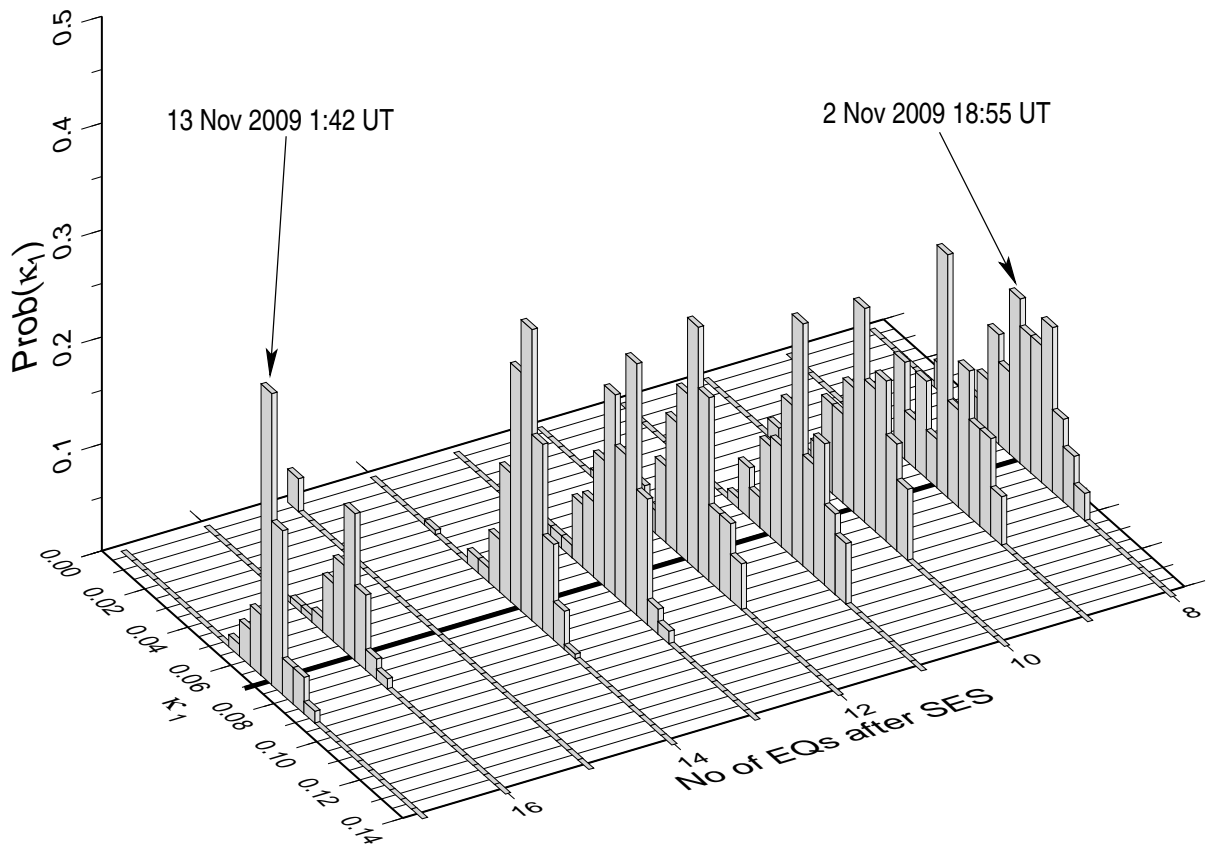}
\caption{The probability Prob($\kappa_1$) of the seismicity within
the area N$_{37.5}^{38.6}$E$_{19.8}^{23.3}$ subsequent to the SES
activity recorded at PAT on October 24, 2009, when considering
M$_{thres}$=3.1. The seismic data until early in the morning on
November 13, 2009, have been used.} \label{fig16}
\end{figure}

{\em Note added on November 14, 2009.} Meanwhile, two SES
activities were recorded at PAT on October 24 and November 11,
2009, see Fig.\ref{fig14}. Hence, the {\em ongoing} study
(Fig.\ref{fig15}) of the previous Note should be accompanied by a
complementary study, focused on the determination of the
occurrence time of the impending mainshock, that also considers
the seismicity in the western part (i.e.,
N$_{37.5}^{38.6}$E$_{19.8}^{22.2}$) of the PAT selectivity map
(see area A in Fig.8 of Ref. \cite{SAR08}). Such a complementary
study, when taking into account the seismic data until early in
the morning on November 13, 2009, leads to the results depicted in
Fig.\ref{fig16} for $M_{thres}$=3.1. A local maximum of
Prob($\kappa_1$) at $\kappa_1=0.070$  was observed (see the
rightmost arrow in Fig.\ref{fig16}) upon the occurrence of
M$_L$=3.2 event at 18:55 UT on November 2, 2009 (that was followed
by a M$_L$=5.6 EQ at 05:25 UT on November 3, 2009, with an
epicenter at $37.4^o$N $20.4^o$E). In addition, the {\em most
prominent} maximum in Fig.\ref{fig16} of Prob($\kappa_1$) at
$\kappa_1=0.070$ is observed later, i.e., upon the occurrence of a
M$_L$=3.1 event at 01:42 UT on November 13, 2009 (see the leftmost
arrow in Fig.\ref{fig16}). This, which has been also checked for
$M_{thres}$=3.0, provides evidence that the critical point is
approached. Calculations are now carried out to further investigate the
magnitude threshold invariance of this behavior.

{\em Note added on November 27, 2009.} Actually, after the
previous Note the following two earthquakes occurred at 18:30 UT
on November 16, 2009 and at 04:29 UT on November 19, 2009 with
magnitudes (M$_s$(ATH)) 4.3 and 4.4 and epicenters at $38.4^o$N
$22.0^o$E and $38.2^o$N $22.7^o$E, respectively, that both lie
within the PAT selectivity map, i.e,
N$_{37.5}^{38.6}$E$_{19.8}^{23.3}$. The continuation of  the two
studies mentioned in the previous Note, i.e., in the PAT
selectivity map as well as in the area
N$_{37.65}^{39.00}$E$_{22.2}^{24.1}$, both show that the {\em most
prominent} maximum of Prob($\kappa_1$) at $\kappa_1=0.070$ was
observed upon the occurrence of the M$_L$=3.2 event at 19:31 UT on
November 25, 2009, thus indicating that the system is near the
critical point.

{\em Note added on February 9, 2010.} Two earthquakes with
magnitudes M$_s$(ATH)($=M_L+0.5$)=5.7 and 5.6 occurred at
$38.4^o$N $22.0^o$E (and hence inside the expected area) at 15:56
UT on January 18 and at 00:46 UT on January 22, 2010 (after
observing a maximum of Prob($\kappa_1$) versus $\kappa_1$ at
$\kappa_1$=0.070 for $M_{thres}$=3.1 upon the occurrence of the
$M_L$=3.2 event at 15:56 UT on January 15, 2010). These two
magnitudes are consistent with the condition M$_s$(ATH)$\geq$6.0
(under which SES information are publicized) if a
reasonable\cite{newbook} experimental error 3$\sigma$=0.7 is
accepted.

\begin{figure}
\includegraphics{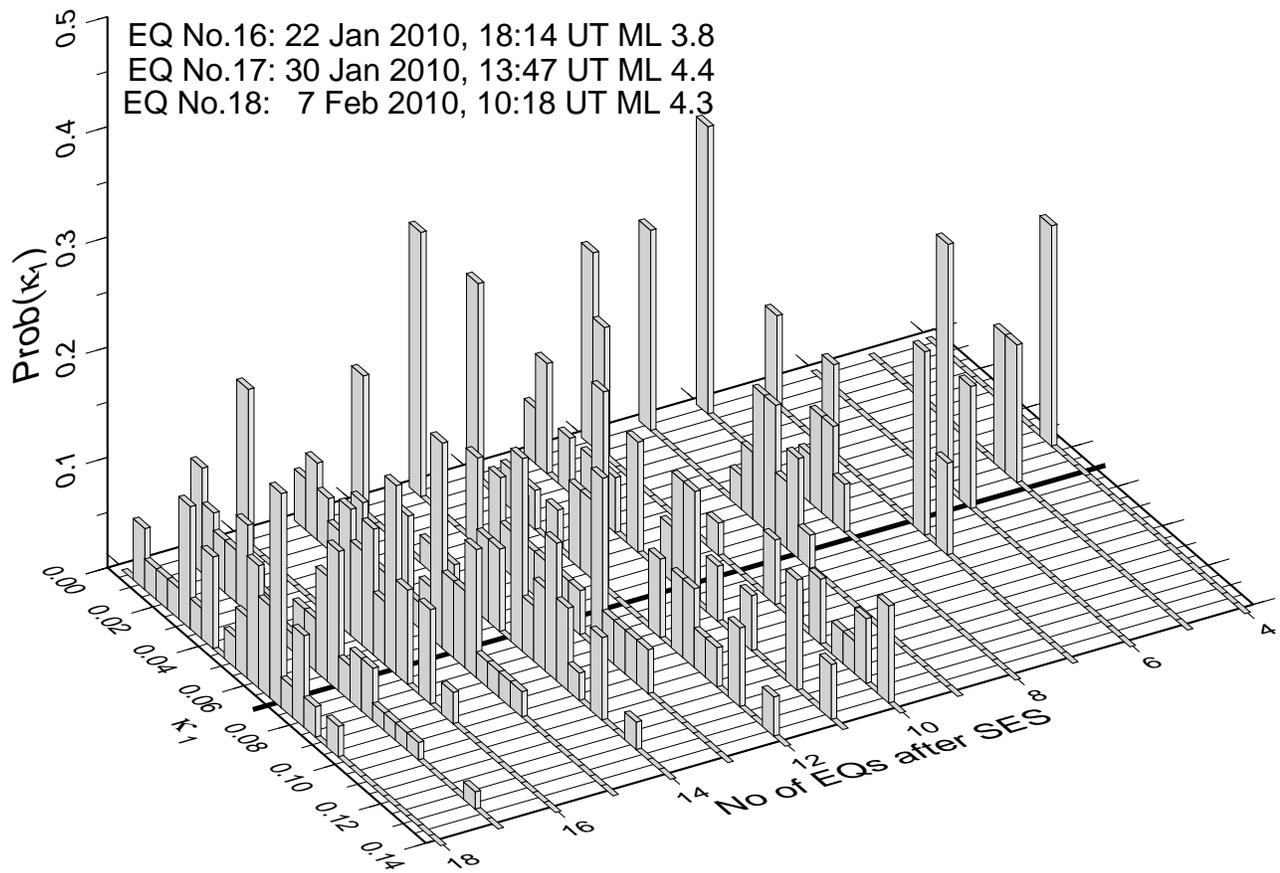}
\caption{The probability Prob($\kappa_1$) of the seismicity, for
M$_{thres}$=3.5, within the area
N$_{38.0}^{39.0}$E$_{21.5}^{23.7}$ when considering the seismic
data until early in the morning on February 8, 2010.}
\label{fig17}
\end{figure}

However, since the amplitudes of all the SES activities recorded
could warrant an impending mainshock with magnitude exceeding
M$_s$(ATH)=6.0, and in order to determine the occurrence time of
such a mainshock, if any, the natural time analysis of seismicity
still continues. This analysis starting at 04:00 UT on December
27, 2009, revealed that in the area
N$_{38.0}^{39.0}$E$_{21.5}^{23.7}$ (almost 200km $\times$ 100km)
the criticality condition (i.e., Prob($\kappa_1$) versus
$\kappa_1$ maximizes at $\kappa_1$=0.070) for $M_{thres}$=3.5 is
obeyed during the last three events of Fig.\ref{fig17}, the latest
of which (that exhibits the strongest maximum) occurred at 10:18
UT on February 7, 2010 at $38.6^o$N $23.7^o$E with $M_L$=4.3. This
has been checked to be valid for broader areas as well (spatial
invariance), but has not yet been found to be confirmed (until
early in the morning on February 8, 2010) for lower magnitude
thresholds. The investigation of the latter is still in progress,
because the following might happen: Due to coarse-graining the
criticality condition might have been obeyed at a larger
time-window for large magnitude thresholds, while for smaller
thresholds the time-window may become smaller thus achieving a
better accuracy in the determination of the occurrence time of the
impending earthquake.

\section{Discussion}
In general, electric field variations are interconnected with the
magnetic field ones through Maxwell equations. Thus, it is
intuitively expected that when the former exhibit long range
correlations the same should hold for the latter. This expectation
is consistent with the present findings which show that, at long
time-lags, the original time-series of both electric- and magnetic
field variations preceding rupture exhibit DFA-exponents close to
unity.

The above can be verified when data of both electric- and
magnetic-field variations are simultaneously available. This was
the case of the data presented in Fig.\ref{fig1m}. In several
occasions, however, as mentioned in Section I, only magnetic field
data exist, in view of the fact that it is easier to conduct
magnetic field measurements compared to the electric field ones.
When using coil magnetometers, the magnetic field variations have
the form of series of `spikes'. Whenever the amplitude of these
`spikes' significantly exceed the pseudo-sinusoidal variations of
the MT background, as in the case of B$_{EW}$ in Fig.\ref{fig1m},
a direct application of DFA (see Fig.\ref{fig1mdfa}) elucidates
the long range correlations in the magnetic field variations
preceding rupture. On the other hand, when considerable
pseudo-sinusoidal MT variations are superimposed, a direct
application of DFA is not advisable. One must first subtract the
MT variations (following a procedure similar to that used in the
electric field data in Fig.\ref{FigFeb}) and then apply DFA.

The preceding paragraph refers to the analysis of the signal in
conventional time. As already shown in Ref.\cite{NAT03B}, natural
time analysis, has the privilege that allows the distinction
between true SES activities and man made signals. This type of
analysis, however, demands the knowledge of the energy released
during each consecutive event. The determination of this energy is
easier to conduct in the case of electric field variations (this
is so, because coil magnetometers, as mentioned in Section II, act
as dB/dt detectors). When these variations are of clear
dichotomous nature, the energy release is proportional to the
duration of each pulse\cite{NAT03,NAT03B}. On the other hand, in
absence of an obvious dichotomous nature an analysis of the
`instantaneous power' similar to that presented in the last
paragraph of Section III should be carried out to determine the
parameters $\kappa_1$, S and S$_-$ in natural time.

\section{Conclusions}
\label{tesera} First, DFA was used as a scaling analysis method to
investigate long-range correlations in the original time series of
the magnetic field variations that precede rupture and have the
form of `spikes' of alternating sign. We find a scaling exponent
$\alpha$ close to 0.9 for time-lags larger than the average time
interval $\langle T \rangle$  between consecutive `spikes', while
at shorter time-lags the exponent is close to 0.5 thus
corresponding to uncorrelated behavior.

Second, using electric field data of long duration SES activities
(i.e., from several hours to a couple of days) recorded during the
last few years, DFA reveals a scale invariant feature with an
exponent $\alpha \approx 1$, over {\em all} scales available
(approximately five orders of magnitude).



\begin{thebibliography}{55}
\expandafter\ifx\csname natexlab\endcsname\relax\def\natexlab#1{#1}\fi
\expandafter\ifx\csname bibnamefont\endcsname\relax
  \def\bibnamefont#1{#1}\fi
\expandafter\ifx\csname bibfnamefont\endcsname\relax
  \def\bibfnamefont#1{#1}\fi
\expandafter\ifx\csname citenamefont\endcsname\relax
  \def\citenamefont#1{#1}\fi
\expandafter\ifx\csname url\endcsname\relax
  \def\url#1{\texttt{#1}}\fi
\expandafter\ifx\csname urlprefix\endcsname\relax\def\urlprefix{URL }\fi
\providecommand{\bibinfo}[2]{#2}
\providecommand{\eprint}[2][]{\url{#2}}

\bibitem[{\citenamefont{Peng et~al.}(1993)\citenamefont{Peng, Mietus,
  Hausdorff, Havlin, Stanley, and Goldberger}}]{PEN93}
\bibinfo{author}{\bibfnamefont{C.-K.} \bibnamefont{Peng}},
  \bibinfo{author}{\bibfnamefont{J.}~\bibnamefont{Mietus}},
  \bibinfo{author}{\bibfnamefont{J.~M.} \bibnamefont{Hausdorff}},
  \bibinfo{author}{\bibfnamefont{S.}~\bibnamefont{Havlin}},
  \bibinfo{author}{\bibfnamefont{H.~E.} \bibnamefont{Stanley}},
  \bibnamefont{and} \bibinfo{author}{\bibfnamefont{A.~L.}
  \bibnamefont{Goldberger}}, \bibinfo{journal}{Phys. Rev. Lett.}
  \textbf{\bibinfo{volume}{70}}, \bibinfo{pages}{1343} (\bibinfo{year}{1993}).

\bibitem[{\citenamefont{Peng et~al.}(1994)\citenamefont{Peng, Buldyrev, Havlin,
  Simons, Stanley, and Goldberger}}]{p18}
\bibinfo{author}{\bibfnamefont{C.-K.} \bibnamefont{Peng}},
  \bibinfo{author}{\bibfnamefont{S.~V.} \bibnamefont{Buldyrev}},
  \bibinfo{author}{\bibfnamefont{S.}~\bibnamefont{Havlin}},
  \bibinfo{author}{\bibfnamefont{M.}~\bibnamefont{Simons}},
  \bibinfo{author}{\bibfnamefont{H.~E.} \bibnamefont{Stanley}},
  \bibnamefont{and} \bibinfo{author}{\bibfnamefont{A.~L.}
  \bibnamefont{Goldberger}}, \bibinfo{journal}{Phys. Rev. E}
  \textbf{\bibinfo{volume}{49}}, \bibinfo{pages}{1685} (\bibinfo{year}{1994}).

\bibitem[{\citenamefont{Buldyrev et~al.}(1995)\citenamefont{Buldyrev,
  Goldberger, Havlin, Mantegna, Matsa, Peng, Simons, and Stanley}}]{p19}
\bibinfo{author}{\bibfnamefont{S.~V.} \bibnamefont{Buldyrev}},
  \bibinfo{author}{\bibfnamefont{A.~L.} \bibnamefont{Goldberger}},
  \bibinfo{author}{\bibfnamefont{S.}~\bibnamefont{Havlin}},
  \bibinfo{author}{\bibfnamefont{R.~N.} \bibnamefont{Mantegna}},
  \bibinfo{author}{\bibfnamefont{M.~E.} \bibnamefont{Matsa}},
  \bibinfo{author}{\bibfnamefont{C.-K.} \bibnamefont{Peng}},
  \bibinfo{author}{\bibfnamefont{M.}~\bibnamefont{Simons}}, \bibnamefont{and}
  \bibinfo{author}{\bibfnamefont{H.~E.} \bibnamefont{Stanley}},
  \bibinfo{journal}{Phys. Rev. E} \textbf{\bibinfo{volume}{51}},
  \bibinfo{pages}{5084} (\bibinfo{year}{1995}).

\bibitem[{\citenamefont{Taqqu et~al.}(1995)\citenamefont{Taqqu, Teverovsky, and
  Willinger}}]{TAQ95}
\bibinfo{author}{\bibfnamefont{M.~S.} \bibnamefont{Taqqu}},
  \bibinfo{author}{\bibfnamefont{V.}~\bibnamefont{Teverovsky}},
  \bibnamefont{and}
  \bibinfo{author}{\bibfnamefont{W.}~\bibnamefont{Willinger}},
  \bibinfo{journal}{Fractals} \textbf{\bibinfo{volume}{3}},
  \bibinfo{pages}{785} (\bibinfo{year}{1995}).

\bibitem[{\citenamefont{Talkner and Weber}(2000)}]{tal00}
\bibinfo{author}{\bibfnamefont{P.}~\bibnamefont{Talkner}} \bibnamefont{and}
  \bibinfo{author}{\bibfnamefont{R.~O.} \bibnamefont{Weber}},
  \bibinfo{journal}{Phys. Rev. E} \textbf{\bibinfo{volume}{62}},
  \bibinfo{pages}{150} (\bibinfo{year}{2000}).

\bibitem[{\citenamefont{Hu et~al.}(2001)\citenamefont{Hu, Ivanov, Chen,
  Carpena, and Stanley}}]{Hu01}
\bibinfo{author}{\bibfnamefont{K.}~\bibnamefont{Hu}},
  \bibinfo{author}{\bibfnamefont{P.~C.} \bibnamefont{Ivanov}},
  \bibinfo{author}{\bibfnamefont{Z.}~\bibnamefont{Chen}},
  \bibinfo{author}{\bibfnamefont{P.}~\bibnamefont{Carpena}}, \bibnamefont{and}
  \bibinfo{author}{\bibfnamefont{H.~E.} \bibnamefont{Stanley}},
  \bibinfo{journal}{Phys. Rev. E} \textbf{\bibinfo{volume}{64}},
  \bibinfo{pages}{011114} (\bibinfo{year}{2001}).

\bibitem[{\citenamefont{Chen et~al.}(2002)\citenamefont{Chen, Ivanov, Hu, and
  Stanley}}]{CHE02}
\bibinfo{author}{\bibfnamefont{Z.}~\bibnamefont{Chen}},
  \bibinfo{author}{\bibfnamefont{P.~C.} \bibnamefont{Ivanov}},
  \bibinfo{author}{\bibfnamefont{K.}~\bibnamefont{Hu}}, \bibnamefont{and}
  \bibinfo{author}{\bibfnamefont{H.~E.} \bibnamefont{Stanley}},
  \bibinfo{journal}{Phys. Rev. E} \textbf{\bibinfo{volume}{65}},
  \bibinfo{pages}{041107} (\bibinfo{year}{2002}).

\bibitem[{\citenamefont{Chen et~al.}(2005)\citenamefont{Chen, Hu, Carpena,
  Bernaola-Galvan, Stanley, and Ivanov}}]{CHE05}
\bibinfo{author}{\bibfnamefont{Z.}~\bibnamefont{Chen}},
  \bibinfo{author}{\bibfnamefont{K.}~\bibnamefont{Hu}},
  \bibinfo{author}{\bibfnamefont{P.}~\bibnamefont{Carpena}},
  \bibinfo{author}{\bibfnamefont{P.}~\bibnamefont{Bernaola-Galvan}},
  \bibinfo{author}{\bibfnamefont{H.~E.} \bibnamefont{Stanley}},
  \bibnamefont{and} \bibinfo{author}{\bibfnamefont{P.~C.}
  \bibnamefont{Ivanov}}, \bibinfo{journal}{Phys. Rev. E}
  \textbf{\bibinfo{volume}{71}}, \bibinfo{pages}{011104}
  (\bibinfo{year}{2005}).

\bibitem[{\citenamefont{Xu et~al.}(2005)\citenamefont{Xu, Ivanov, Hu, Chen,
  Carbone, and Stanley}}]{Hu05}
\bibinfo{author}{\bibfnamefont{L.}~\bibnamefont{Xu}},
  \bibinfo{author}{\bibfnamefont{P.~C.} \bibnamefont{Ivanov}},
  \bibinfo{author}{\bibfnamefont{K.}~\bibnamefont{Hu}},
  \bibinfo{author}{\bibfnamefont{Z.}~\bibnamefont{Chen}},
  \bibinfo{author}{\bibfnamefont{A.}~\bibnamefont{Carbone}}, \bibnamefont{and}
  \bibinfo{author}{\bibfnamefont{H.~E.} \bibnamefont{Stanley}},
  \bibinfo{journal}{Phys. Rev. E} \textbf{\bibinfo{volume}{71}},
  \bibinfo{pages}{051101} (\bibinfo{year}{2005}).

\bibitem[{\citenamefont{Ossadnik et~al.}(1994)\citenamefont{Ossadnik, Buldyrev,
  Goldberger, Havlin, Mantegna, Peng, Simons, and Stanley}}]{OSS94}
\bibinfo{author}{\bibfnamefont{S.~M.} \bibnamefont{Ossadnik}},
  \bibinfo{author}{\bibfnamefont{S.~B.} \bibnamefont{Buldyrev}},
  \bibinfo{author}{\bibfnamefont{A.~L.} \bibnamefont{Goldberger}},
  \bibinfo{author}{\bibfnamefont{S.}~\bibnamefont{Havlin}},
  \bibinfo{author}{\bibfnamefont{R.~N.} \bibnamefont{Mantegna}},
  \bibinfo{author}{\bibfnamefont{C.~K.} \bibnamefont{Peng}},
  \bibinfo{author}{\bibfnamefont{M.}~\bibnamefont{Simons}}, \bibnamefont{and}
  \bibinfo{author}{\bibfnamefont{H.~E.} \bibnamefont{Stanley}},
  \bibinfo{journal}{Biophys. J.} \textbf{\bibinfo{volume}{67}},
  \bibinfo{pages}{64} (\bibinfo{year}{1994}).

\bibitem[{\citenamefont{Mantegna et~al.}(1994)\citenamefont{Mantegna, Buldyrev,
  Goldberger, Havlin, , Peng, Simons, and Stanley}}]{MAN94}
\bibinfo{author}{\bibfnamefont{R.~N.} \bibnamefont{Mantegna}},
  \bibinfo{author}{\bibfnamefont{S.~V.} \bibnamefont{Buldyrev}},
  \bibinfo{author}{\bibfnamefont{A.~L.} \bibnamefont{Goldberger}},
  \bibinfo{author}{\bibfnamefont{S.}~\bibnamefont{Havlin}}, ,
  \bibinfo{author}{\bibfnamefont{C.~K.} \bibnamefont{Peng}},
  \bibinfo{author}{\bibfnamefont{M.}~\bibnamefont{Simons}}, \bibnamefont{and}
  \bibinfo{author}{\bibfnamefont{H.~E.} \bibnamefont{Stanley}},
  \bibinfo{journal}{Phys. Rev. Lett.} \textbf{\bibinfo{volume}{73}},
  \bibinfo{pages}{3169} (\bibinfo{year}{1994}).

\bibitem[{\citenamefont{Mantegna et~al.}(1996)\citenamefont{Mantegna, Buldyrev,
  Goldberger, Havlin, , Peng, Simons, and Stanley}}]{MAN96}
\bibinfo{author}{\bibfnamefont{R.~N.} \bibnamefont{Mantegna}},
  \bibinfo{author}{\bibfnamefont{S.~V.} \bibnamefont{Buldyrev}},
  \bibinfo{author}{\bibfnamefont{A.~L.} \bibnamefont{Goldberger}},
  \bibinfo{author}{\bibfnamefont{S.}~\bibnamefont{Havlin}}, ,
  \bibinfo{author}{\bibfnamefont{C.~K.} \bibnamefont{Peng}},
  \bibinfo{author}{\bibfnamefont{M.}~\bibnamefont{Simons}}, \bibnamefont{and}
  \bibinfo{author}{\bibfnamefont{H.~E.} \bibnamefont{Stanley}},
  \bibinfo{journal}{Phys. Rev. Lett.} \textbf{\bibinfo{volume}{76}},
  \bibinfo{pages}{1979} (\bibinfo{year}{1996}).

\bibitem[{\citenamefont{Carpena et~al.}(2002)\citenamefont{Carpena,
  Bernaola-Galv\'{a}n, Ivanov, and Stanley}}]{CAR02}
\bibinfo{author}{\bibfnamefont{P.}~\bibnamefont{Carpena}},
  \bibinfo{author}{\bibfnamefont{P.}~\bibnamefont{Bernaola-Galv\'{a}n}},
  \bibinfo{author}{\bibfnamefont{P.~C.} \bibnamefont{Ivanov}},
  \bibnamefont{and} \bibinfo{author}{\bibfnamefont{H.~E.}
  \bibnamefont{Stanley}}, \bibinfo{journal}{Nature (London)}
  \textbf{\bibinfo{volume}{418}}, \bibinfo{pages}{955} (\bibinfo{year}{2002}).

\bibitem[{\citenamefont{Hu et~al.}(2004)\citenamefont{Hu, Ivanov, Chen, Hilton,
  Stanley, and Shea}}]{HU04}
\bibinfo{author}{\bibfnamefont{K.}~\bibnamefont{Hu}},
  \bibinfo{author}{\bibfnamefont{P.~C.} \bibnamefont{Ivanov}},
  \bibinfo{author}{\bibfnamefont{Z.}~\bibnamefont{Chen}},
  \bibinfo{author}{\bibfnamefont{M.~F.} \bibnamefont{Hilton}},
  \bibinfo{author}{\bibfnamefont{H.~E.} \bibnamefont{Stanley}},
  \bibnamefont{and} \bibinfo{author}{\bibfnamefont{S.~A.} \bibnamefont{Shea}},
  \bibinfo{journal}{Physica A} \textbf{\bibinfo{volume}{337}},
  \bibinfo{pages}{307} (\bibinfo{year}{2004}).

\bibitem[{\citenamefont{Hausdorff et~al.}(2001)\citenamefont{Hausdorff,
  Ashkenazy, Peng, Ivanov, Stanley, and Goldberger}}]{HAUS01}
\bibinfo{author}{\bibfnamefont{J.~M.} \bibnamefont{Hausdorff}},
  \bibinfo{author}{\bibfnamefont{Y.}~\bibnamefont{Ashkenazy}},
  \bibinfo{author}{\bibfnamefont{C.~K.} \bibnamefont{Peng}},
  \bibinfo{author}{\bibfnamefont{P.~C.} \bibnamefont{Ivanov}},
  \bibinfo{author}{\bibfnamefont{H.~E.} \bibnamefont{Stanley}},
  \bibnamefont{and} \bibinfo{author}{\bibfnamefont{A.~L.}
  \bibnamefont{Goldberger}}, \bibinfo{journal}{Physica A}
  \textbf{\bibinfo{volume}{302}}, \bibinfo{pages}{138} (\bibinfo{year}{2001}).

\bibitem[{\citenamefont{Ashkenazy et~al.}(2002)\citenamefont{Ashkenazy,
  Hausdorff, Ivanov, and Stanley}}]{ASH02}
\bibinfo{author}{\bibfnamefont{Y.}~\bibnamefont{Ashkenazy}},
  \bibinfo{author}{\bibfnamefont{J.~M.} \bibnamefont{Hausdorff}},
  \bibinfo{author}{\bibfnamefont{P.~C.} \bibnamefont{Ivanov}},
  \bibnamefont{and} \bibinfo{author}{\bibfnamefont{H.~E.}
  \bibnamefont{Stanley}}, \bibinfo{journal}{Physica A}
  \textbf{\bibinfo{volume}{316}}, \bibinfo{pages}{662} (\bibinfo{year}{2002}).

\bibitem[{\citenamefont{Ivanov et~al.}(1999)\citenamefont{Ivanov, Bunde,
  Nunes~Amaral, Havlin, Fritsch-Yelle, Baevsky, Stanley, and
  Goldberger}}]{IVA99}
\bibinfo{author}{\bibfnamefont{P.~C.} \bibnamefont{Ivanov}},
  \bibinfo{author}{\bibfnamefont{A.}~\bibnamefont{Bunde}},
  \bibinfo{author}{\bibfnamefont{L.~A.} \bibnamefont{Nunes~Amaral}},
  \bibinfo{author}{\bibfnamefont{S.}~\bibnamefont{Havlin}},
  \bibinfo{author}{\bibfnamefont{J.}~\bibnamefont{Fritsch-Yelle}},
  \bibinfo{author}{\bibfnamefont{R.~M.} \bibnamefont{Baevsky}},
  \bibinfo{author}{\bibfnamefont{H.~E.} \bibnamefont{Stanley}},
  \bibnamefont{and} \bibinfo{author}{\bibfnamefont{A.~L.}
  \bibnamefont{Goldberger}}, \bibinfo{journal}{Europhys. Lett.}
  \textbf{\bibinfo{volume}{48}}, \bibinfo{pages}{594} (\bibinfo{year}{1999}).

\bibitem[{\citenamefont{Havlin et~al.}(1999)\citenamefont{Havlin, Buldyrev,
  Bunde, Goldberger, Ivanov, Peng, and Stanley}}]{HAV99}
\bibinfo{author}{\bibfnamefont{S.}~\bibnamefont{Havlin}},
  \bibinfo{author}{\bibfnamefont{S.~V.} \bibnamefont{Buldyrev}},
  \bibinfo{author}{\bibfnamefont{A.}~\bibnamefont{Bunde}},
  \bibinfo{author}{\bibfnamefont{A.~L.} \bibnamefont{Goldberger}},
  \bibinfo{author}{\bibfnamefont{P.~C.} \bibnamefont{Ivanov}},
  \bibinfo{author}{\bibfnamefont{C.~K.} \bibnamefont{Peng}}, \bibnamefont{and}
  \bibinfo{author}{\bibfnamefont{H.~E.} \bibnamefont{Stanley}},
  \bibinfo{journal}{Physica A} \textbf{\bibinfo{volume}{273}},
  \bibinfo{pages}{46} (\bibinfo{year}{1999}).

\bibitem[{\citenamefont{Stanley et~al.}(1999)\citenamefont{Stanley,
  Nunes~Amaral, Goldberger, Havlin, Ivanov, and Peng}}]{STAN99}
\bibinfo{author}{\bibfnamefont{H.~E.} \bibnamefont{Stanley}},
  \bibinfo{author}{\bibfnamefont{L.~A.} \bibnamefont{Nunes~Amaral}},
  \bibinfo{author}{\bibfnamefont{A.~L.} \bibnamefont{Goldberger}},
  \bibinfo{author}{\bibfnamefont{S.}~\bibnamefont{Havlin}},
  \bibinfo{author}{\bibfnamefont{P.~C.} \bibnamefont{Ivanov}},
  \bibnamefont{and} \bibinfo{author}{\bibfnamefont{C.~K.} \bibnamefont{Peng}},
  \bibinfo{journal}{Physica A} \textbf{\bibinfo{volume}{270}},
  \bibinfo{pages}{309} (\bibinfo{year}{1999}).

\bibitem[{\citenamefont{Ivanov et~al.}(1998)\citenamefont{Ivanov, Nunes~Amaral,
  Goldberger, and Stanley}}]{IVA98}
\bibinfo{author}{\bibfnamefont{P.~C.} \bibnamefont{Ivanov}},
  \bibinfo{author}{\bibfnamefont{L.~A.} \bibnamefont{Nunes~Amaral}},
  \bibinfo{author}{\bibfnamefont{A.~L.} \bibnamefont{Goldberger}},
  \bibnamefont{and} \bibinfo{author}{\bibfnamefont{H.~E.}
  \bibnamefont{Stanley}}, \bibinfo{journal}{Europhys. Lett.}
  \textbf{\bibinfo{volume}{43}}, \bibinfo{pages}{363} (\bibinfo{year}{1998}).

\bibitem[{\citenamefont{Ivanova and Ausloos}(1999)}]{IVANOVA99}
\bibinfo{author}{\bibfnamefont{K.}~\bibnamefont{Ivanova}} \bibnamefont{and}
  \bibinfo{author}{\bibfnamefont{M.}~\bibnamefont{Ausloos}},
  \bibinfo{journal}{Physica A} \textbf{\bibinfo{volume}{274}},
  \bibinfo{pages}{349} (\bibinfo{year}{1999}).

\bibitem[{\citenamefont{Ivanova et~al.}(2003)\citenamefont{Ivanova, Ackerman,
  Clothiaux, Ivanov, Stanley, and Ausloos}}]{IVANOVA03}
\bibinfo{author}{\bibfnamefont{K.}~\bibnamefont{Ivanova}},
  \bibinfo{author}{\bibfnamefont{T.~P.} \bibnamefont{Ackerman}},
  \bibinfo{author}{\bibfnamefont{E.~E.} \bibnamefont{Clothiaux}},
  \bibinfo{author}{\bibfnamefont{P.~C.} \bibnamefont{Ivanov}},
  \bibinfo{author}{\bibfnamefont{H.~E.} \bibnamefont{Stanley}},
  \bibnamefont{and} \bibinfo{author}{\bibfnamefont{M.}~\bibnamefont{Ausloos}},
  \bibinfo{journal}{J. Geophys. Res.-Atmospheres} \textbf{\bibinfo{volume}{108
  D9}}, \bibinfo{pages}{4268} (\bibinfo{year}{2003}).

\bibitem[{\citenamefont{Koscielny-Bunde
  et~al.}(1998)\citenamefont{Koscielny-Bunde, Bunde, Havlin, Roman, Goldreich,
  and Schellnhuber}}]{KOS98}
\bibinfo{author}{\bibfnamefont{E.}~\bibnamefont{Koscielny-Bunde}},
  \bibinfo{author}{\bibfnamefont{A.}~\bibnamefont{Bunde}},
  \bibinfo{author}{\bibfnamefont{S.}~\bibnamefont{Havlin}},
  \bibinfo{author}{\bibfnamefont{H.~E.} \bibnamefont{Roman}},
  \bibinfo{author}{\bibfnamefont{Y.}~\bibnamefont{Goldreich}},
  \bibnamefont{and} \bibinfo{author}{\bibfnamefont{H.~J.}
  \bibnamefont{Schellnhuber}}, \bibinfo{journal}{Phys. Rev. Lett.}
  \textbf{\bibinfo{volume}{81}}, \bibinfo{pages}{729} (\bibinfo{year}{1998}).

\bibitem[{\citenamefont{Stratonovich}(1981)}]{Stra81}
\bibinfo{author}{\bibfnamefont{R.~L.} \bibnamefont{Stratonovich}},
  \emph{\bibinfo{title}{Topics in the theory of random noise Vol. I}}
  (\bibinfo{publisher}{Gordon and Breach}, \bibinfo{address}{New York},
  \bibinfo{year}{1981}).

\bibitem[{\citenamefont{Varotsos et~al.}(2002)\citenamefont{Varotsos, Sarlis,
  and Skordas}}]{NAT02}
\bibinfo{author}{\bibfnamefont{P.~A.} \bibnamefont{Varotsos}},
  \bibinfo{author}{\bibfnamefont{N.~V.} \bibnamefont{Sarlis}},
  \bibnamefont{and} \bibinfo{author}{\bibfnamefont{E.~S.}
  \bibnamefont{Skordas}}, \bibinfo{journal}{Phys. Rev. E}
  \textbf{\bibinfo{volume}{66}}, \bibinfo{pages}{011902}
  (\bibinfo{year}{2002}).

\bibitem[{\citenamefont{Weron et~al.}(2005)\citenamefont{Weron, Burnecki,
  Mercik, and Weron}}]{WER05}
\bibinfo{author}{\bibfnamefont{A.}~\bibnamefont{Weron}},
  \bibinfo{author}{\bibfnamefont{K.}~\bibnamefont{Burnecki}},
  \bibinfo{author}{\bibfnamefont{S.}~\bibnamefont{Mercik}}, \bibnamefont{and}
  \bibinfo{author}{\bibfnamefont{K.}~\bibnamefont{Weron}},
  \bibinfo{journal}{Phys. Rev. E} \textbf{\bibinfo{volume}{71}},
  \bibinfo{pages}{016113} (\bibinfo{year}{2005}).

\bibitem[{\citenamefont{Varotsos and Alexopoulos}(1984)}]{proto}
\bibinfo{author}{\bibfnamefont{P.}~\bibnamefont{Varotsos}} \bibnamefont{and}
  \bibinfo{author}{\bibfnamefont{K.}~\bibnamefont{Alexopoulos}},
  \bibinfo{journal}{Tectonophysics} \textbf{\bibinfo{volume}{110}},
  \bibinfo{pages}{73} (\bibinfo{year}{1984}).

\bibitem[{\citenamefont{Fraser-Smith et~al.}(1990)\citenamefont{Fraser-Smith,
  Bernardi, McGill, Ladd, Helliwell, and Villard}}]{FRA90}
\bibinfo{author}{\bibfnamefont{A.~C.} \bibnamefont{Fraser-Smith}},
  \bibinfo{author}{\bibfnamefont{A.}~\bibnamefont{Bernardi}},
  \bibinfo{author}{\bibfnamefont{P.~R.} \bibnamefont{McGill}},
  \bibinfo{author}{\bibfnamefont{M.~E.} \bibnamefont{Ladd}},
  \bibinfo{author}{\bibfnamefont{R.~A.} \bibnamefont{Helliwell}},
  \bibnamefont{and} \bibinfo{author}{\bibfnamefont{O.~G.}
  \bibnamefont{Villard}}, \bibinfo{journal}{Geophys. Res. Lett.}
  \textbf{\bibinfo{volume}{17}}, \bibinfo{pages}{1465} (\bibinfo{year}{1990}).

\bibitem[{\citenamefont{Uyeda et~al.}(2000)\citenamefont{Uyeda, Nagao, Orihara,
  Yamaguchi, and Takahashi}}]{uye}
\bibinfo{author}{\bibfnamefont{S.}~\bibnamefont{Uyeda}},
  \bibinfo{author}{\bibfnamefont{T.}~\bibnamefont{Nagao}},
  \bibinfo{author}{\bibfnamefont{Y.}~\bibnamefont{Orihara}},
  \bibinfo{author}{\bibfnamefont{T.}~\bibnamefont{Yamaguchi}},
  \bibnamefont{and}
  \bibinfo{author}{\bibfnamefont{I.}~\bibnamefont{Takahashi}},
  \bibinfo{journal}{Proc. Natl. Acad. Sci. USA} \textbf{\bibinfo{volume}{97}},
  \bibinfo{pages}{4561} (\bibinfo{year}{2000}).

\bibitem[{\citenamefont{Uyeda et~al.}(2002)\citenamefont{Uyeda, Hayakawa,
  Nagao, Molchanov, Hattori, Orihara, Gotoh, Akinaga, and Tanaka}}]{uye2}
\bibinfo{author}{\bibfnamefont{S.}~\bibnamefont{Uyeda}},
  \bibinfo{author}{\bibfnamefont{M.}~\bibnamefont{Hayakawa}},
  \bibinfo{author}{\bibfnamefont{T.}~\bibnamefont{Nagao}},
  \bibinfo{author}{\bibfnamefont{O.}~\bibnamefont{Molchanov}},
  \bibinfo{author}{\bibfnamefont{K.}~\bibnamefont{Hattori}},
  \bibinfo{author}{\bibfnamefont{Y.}~\bibnamefont{Orihara}},
  \bibinfo{author}{\bibfnamefont{K.}~\bibnamefont{Gotoh}},
  \bibinfo{author}{\bibfnamefont{Y.}~\bibnamefont{Akinaga}}, \bibnamefont{and}
  \bibinfo{author}{\bibfnamefont{H.}~\bibnamefont{Tanaka}},
  \bibinfo{journal}{Proc. Natl. Acad. Sci. USA} \textbf{\bibinfo{volume}{99}},
  \bibinfo{pages}{7352} (\bibinfo{year}{2002}).

\bibitem[{\citenamefont{Varotsos
  et~al.}(2005{\natexlab{a}})\citenamefont{Varotsos, Sarlis, Tanaka, and
  Skordas}}]{VAR05C}
\bibinfo{author}{\bibfnamefont{P.~A.} \bibnamefont{Varotsos}},
  \bibinfo{author}{\bibfnamefont{N.~V.} \bibnamefont{Sarlis}},
  \bibinfo{author}{\bibfnamefont{H.~K.} \bibnamefont{Tanaka}},
  \bibnamefont{and} \bibinfo{author}{\bibfnamefont{E.~S.}
  \bibnamefont{Skordas}}, \bibinfo{journal}{Phys. Rev. E}
  \textbf{\bibinfo{volume}{72}}, \bibinfo{pages}{041103}
  (\bibinfo{year}{2005}{\natexlab{a}}).

\bibitem[{\citenamefont{Sarlis et~al.}(2008)\citenamefont{Sarlis, Skordas,
  Lazaridou, and Varotsos}}]{SAR08}
\bibinfo{author}{\bibfnamefont{N.~V.} \bibnamefont{Sarlis}},
  \bibinfo{author}{\bibfnamefont{E.~S.} \bibnamefont{Skordas}},
  \bibinfo{author}{\bibfnamefont{M.~S.} \bibnamefont{Lazaridou}},
  \bibnamefont{and} \bibinfo{author}{\bibfnamefont{P.~A.}
  \bibnamefont{Varotsos}}, \bibinfo{journal}{Proc. Japan Acad., Ser. B}
  \textbf{\bibinfo{volume}{84}}, \bibinfo{pages}{331} (\bibinfo{year}{2008}).

\bibitem[{\citenamefont{Varotsos et~al.}(1982)\citenamefont{Varotsos,
  Alexopoulos, and Nomicos}}]{VARALEX82}
\bibinfo{author}{\bibfnamefont{P.}~\bibnamefont{Varotsos}},
  \bibinfo{author}{\bibfnamefont{K.}~\bibnamefont{Alexopoulos}},
  \bibnamefont{and} \bibinfo{author}{\bibfnamefont{K.}~\bibnamefont{Nomicos}},
  \bibinfo{journal}{Phys. Status Solidi B} \textbf{\bibinfo{volume}{111}},
  \bibinfo{pages}{581} (\bibinfo{year}{1982}).

\bibitem[{\citenamefont{Kostopoulos et~al.}(1975)\citenamefont{Kostopoulos,
  Varotsos, and Mourikis}}]{KOS75}
\bibinfo{author}{\bibfnamefont{D.}~\bibnamefont{Kostopoulos}},
  \bibinfo{author}{\bibfnamefont{P.}~\bibnamefont{Varotsos}}, \bibnamefont{and}
  \bibinfo{author}{\bibfnamefont{S.}~\bibnamefont{Mourikis}},
  \bibinfo{journal}{Can. J. Phys.} \textbf{\bibinfo{volume}{53}},
  \bibinfo{pages}{1318} (\bibinfo{year}{1975}).

\bibitem[{\citenamefont{Varotsos}(1981)}]{VAR81}
\bibinfo{author}{\bibfnamefont{P.}~\bibnamefont{Varotsos}},
  \bibinfo{journal}{J. Phys. Chem. Sol.} \textbf{\bibinfo{volume}{42}},
  \bibinfo{pages}{405} (\bibinfo{year}{1981}).

\bibitem[{\citenamefont{Varotsos and
  Alexopoulos}(1980{\natexlab{a}})}]{VARALEX80}
\bibinfo{author}{\bibfnamefont{P.}~\bibnamefont{Varotsos}} \bibnamefont{and}
  \bibinfo{author}{\bibfnamefont{K.}~\bibnamefont{Alexopoulos}},
  \bibinfo{journal}{J. Phys. Chem. Sol.} \textbf{\bibinfo{volume}{41}},
  \bibinfo{pages}{443} (\bibinfo{year}{1980}{\natexlab{a}}).

\bibitem[{\citenamefont{Varotsos and Alexopoulos}(1981)}]{VARALEX81}
\bibinfo{author}{\bibfnamefont{P.}~\bibnamefont{Varotsos}} \bibnamefont{and}
  \bibinfo{author}{\bibfnamefont{K.}~\bibnamefont{Alexopoulos}},
  \bibinfo{journal}{Phys. Status Solidi B} \textbf{\bibinfo{volume}{42}},
  \bibinfo{pages}{409} (\bibinfo{year}{1981}).

\bibitem[{\citenamefont{Varotsos and
  Alexopoulos}(1980{\natexlab{b}})}]{VARALEX80p}
\bibinfo{author}{\bibfnamefont{P.}~\bibnamefont{Varotsos}} \bibnamefont{and}
  \bibinfo{author}{\bibfnamefont{K.}~\bibnamefont{Alexopoulos}},
  \bibinfo{journal}{Phys. Rev. B} \textbf{\bibinfo{volume}{21}},
  \bibinfo{pages}{4898} (\bibinfo{year}{1980}{\natexlab{b}}).

\bibitem[{\citenamefont{Varotsos and Alexopoulos}(1986)}]{varbook}
\bibinfo{author}{\bibfnamefont{P.}~\bibnamefont{Varotsos}} \bibnamefont{and}
  \bibinfo{author}{\bibfnamefont{K.}~\bibnamefont{Alexopoulos}},
  \emph{\bibinfo{title}{Thermodynamics of Point Defects and their Relation with
  Bulk Properties}} (\bibinfo{publisher}{North Holland},
  \bibinfo{address}{Amsterdam}, \bibinfo{year}{1986}).

\bibitem[{\citenamefont{Varotsos
  et~al.}(2003{\natexlab{a}})\citenamefont{Varotsos, Sarlis, and
  Skordas}}]{NAT03}
\bibinfo{author}{\bibfnamefont{P.~A.} \bibnamefont{Varotsos}},
  \bibinfo{author}{\bibfnamefont{N.~V.} \bibnamefont{Sarlis}},
  \bibnamefont{and} \bibinfo{author}{\bibfnamefont{E.~S.}
  \bibnamefont{Skordas}}, \bibinfo{journal}{Phys. Rev. E}
  \textbf{\bibinfo{volume}{67}}, \bibinfo{pages}{021109}
  (\bibinfo{year}{2003}{\natexlab{a}}).

\bibitem[{\citenamefont{Varotsos
  et~al.}(2003{\natexlab{b}})\citenamefont{Varotsos, Sarlis, and
  Skordas}}]{NAT03B}
\bibinfo{author}{\bibfnamefont{P.~A.} \bibnamefont{Varotsos}},
  \bibinfo{author}{\bibfnamefont{N.~V.} \bibnamefont{Sarlis}},
  \bibnamefont{and} \bibinfo{author}{\bibfnamefont{E.~S.}
  \bibnamefont{Skordas}}, \bibinfo{journal}{Phys. Rev. E}
  \textbf{\bibinfo{volume}{68}}, \bibinfo{pages}{031106}
  (\bibinfo{year}{2003}{\natexlab{b}}).

\bibitem[{\citenamefont{Varotsos et~al.}(2004)\citenamefont{Varotsos, Sarlis,
  Skordas, and Lazaridou}}]{NAT04}
\bibinfo{author}{\bibfnamefont{P.~A.} \bibnamefont{Varotsos}},
  \bibinfo{author}{\bibfnamefont{N.~V.} \bibnamefont{Sarlis}},
  \bibinfo{author}{\bibfnamefont{E.~S.} \bibnamefont{Skordas}},
  \bibnamefont{and} \bibinfo{author}{\bibfnamefont{M.~S.}
  \bibnamefont{Lazaridou}}, \bibinfo{journal}{Phys. Rev. E}
  \textbf{\bibinfo{volume}{70}}, \bibinfo{pages}{011106}
  (\bibinfo{year}{2004}).

\bibitem[{\citenamefont{Varotsos
  et~al.}(2005{\natexlab{b}})\citenamefont{Varotsos, Sarlis, Tanaka, and
  Skordas}}]{NAT05B}
\bibinfo{author}{\bibfnamefont{P.~A.} \bibnamefont{Varotsos}},
  \bibinfo{author}{\bibfnamefont{N.~V.} \bibnamefont{Sarlis}},
  \bibinfo{author}{\bibfnamefont{H.~K.} \bibnamefont{Tanaka}},
  \bibnamefont{and} \bibinfo{author}{\bibfnamefont{E.~S.}
  \bibnamefont{Skordas}}, \bibinfo{journal}{Phys. Rev. E}
  \textbf{\bibinfo{volume}{71}}, \bibinfo{pages}{032102}
  (\bibinfo{year}{2005}{\natexlab{b}}).

\bibitem[{\citenamefont{Varotsos
  et~al.}(2006{\natexlab{a}})\citenamefont{Varotsos, Sarlis, Skordas, Tanaka,
  and Lazaridou}}]{NAT06A}
\bibinfo{author}{\bibfnamefont{P.~A.} \bibnamefont{Varotsos}},
  \bibinfo{author}{\bibfnamefont{N.~V.} \bibnamefont{Sarlis}},
  \bibinfo{author}{\bibfnamefont{E.~S.} \bibnamefont{Skordas}},
  \bibinfo{author}{\bibfnamefont{H.~K.} \bibnamefont{Tanaka}},
  \bibnamefont{and} \bibinfo{author}{\bibfnamefont{M.~S.}
  \bibnamefont{Lazaridou}}, \bibinfo{journal}{Phys. Rev. E}
  \textbf{\bibinfo{volume}{73}}, \bibinfo{pages}{031114}
  (\bibinfo{year}{2006}{\natexlab{a}}).

\bibitem[{\citenamefont{Lesche}(1982)}]{LES82}
\bibinfo{author}{\bibfnamefont{B.}~\bibnamefont{Lesche}}, \bibinfo{journal}{J.
  Stat. Phys.} \textbf{\bibinfo{volume}{27}}, \bibinfo{pages}{419}
  (\bibinfo{year}{1982}).

\bibitem[{\citenamefont{Lesche}(2004)}]{LES04}
\bibinfo{author}{\bibfnamefont{B.}~\bibnamefont{Lesche}},
  \bibinfo{journal}{Phys. Rev. E} \textbf{\bibinfo{volume}{70}},
  \bibinfo{pages}{017102} (\bibinfo{year}{2004}).

\bibitem[{\citenamefont{Varotsos
  et~al.}(2003{\natexlab{c}})\citenamefont{Varotsos, Sarlis, and
  Skordas}}]{PRL03}
\bibinfo{author}{\bibfnamefont{P.~A.} \bibnamefont{Varotsos}},
  \bibinfo{author}{\bibfnamefont{N.~V.} \bibnamefont{Sarlis}},
  \bibnamefont{and} \bibinfo{author}{\bibfnamefont{E.~S.}
  \bibnamefont{Skordas}}, \bibinfo{journal}{Phys. Rev. Lett.}
  \textbf{\bibinfo{volume}{91}}, \bibinfo{pages}{148501}
  (\bibinfo{year}{2003}{\natexlab{c}}).

\bibitem[{\citenamefont{Sarlis and Varotsos}(2002)}]{sar02}
\bibinfo{author}{\bibfnamefont{N.}~\bibnamefont{Sarlis}} \bibnamefont{and}
  \bibinfo{author}{\bibfnamefont{P.}~\bibnamefont{Varotsos}},
  \bibinfo{journal}{J. Geodyn.} \textbf{\bibinfo{volume}{33}},
  \bibinfo{pages}{463} (\bibinfo{year}{2002}).

\bibitem[{\citenamefont{Karakelian et~al.}(2002)\citenamefont{Karakelian,
  Beroza, Klemperer, and Fraser-Smith}}]{KAR02B}
\bibinfo{author}{\bibfnamefont{D.}~\bibnamefont{Karakelian}},
  \bibinfo{author}{\bibfnamefont{G.~C.} \bibnamefont{Beroza}},
  \bibinfo{author}{\bibfnamefont{S.~L.} \bibnamefont{Klemperer}},
  \bibnamefont{and} \bibinfo{author}{\bibfnamefont{A.~C.}
  \bibnamefont{Fraser-Smith}}, \bibinfo{journal}{Bull. Seism. Soc. Am.}
  \textbf{\bibinfo{volume}{92}}, \bibinfo{pages}{1513} (\bibinfo{year}{2002}).

\bibitem[{\citenamefont{Molchanov et~al.}(1992)\citenamefont{Molchanov,
  Kopytenko, Voronov, Kopytenko, Matiashvili, Fraser-Smith, and
  Bernardi}}]{MOL92}
\bibinfo{author}{\bibfnamefont{O.~A.} \bibnamefont{Molchanov}},
  \bibinfo{author}{\bibfnamefont{Y.~A.} \bibnamefont{Kopytenko}},
  \bibinfo{author}{\bibfnamefont{P.~M.} \bibnamefont{Voronov}},
  \bibinfo{author}{\bibfnamefont{E.~A.} \bibnamefont{Kopytenko}},
  \bibinfo{author}{\bibfnamefont{T.~G.} \bibnamefont{Matiashvili}},
  \bibinfo{author}{\bibfnamefont{A.~C.} \bibnamefont{Fraser-Smith}},
  \bibnamefont{and} \bibinfo{author}{\bibfnamefont{A.}~\bibnamefont{Bernardi}},
  \bibinfo{journal}{Geophys. Res. Lett.} \textbf{\bibinfo{volume}{19}},
  \bibinfo{pages}{1495} (\bibinfo{year}{1992}).

\bibitem[{\citenamefont{Varotsos et~al.}(2001)\citenamefont{Varotsos, Sarlis,
  and Skordas}}]{PJA}
\bibinfo{author}{\bibfnamefont{P.}~\bibnamefont{Varotsos}},
  \bibinfo{author}{\bibfnamefont{N.}~\bibnamefont{Sarlis}}, \bibnamefont{and}
  \bibinfo{author}{\bibfnamefont{E.}~\bibnamefont{Skordas}},
  \bibinfo{journal}{Proc. Jpn. Acad., Ser. B: Phys. Biol. Sci.}
  \textbf{\bibinfo{volume}{77}}, \bibinfo{pages}{87} (\bibinfo{year}{2001}).

\bibitem[{\citenamefont{Varotsos
  et~al.}(2005{\natexlab{c}})\citenamefont{Varotsos, Sarlis, and
  Skordas}}]{APL05}
\bibinfo{author}{\bibfnamefont{P.~A.} \bibnamefont{Varotsos}},
  \bibinfo{author}{\bibfnamefont{N.~V.} \bibnamefont{Sarlis}},
  \bibnamefont{and} \bibinfo{author}{\bibfnamefont{E.~S.}
  \bibnamefont{Skordas}}, \bibinfo{journal}{Appl. Phys. Lett.}
  \textbf{\bibinfo{volume}{86}}, \bibinfo{pages}{194101}
  (\bibinfo{year}{2005}{\natexlab{c}}).

\bibitem[{\citenamefont{Varotsos
  et~al.}(2006{\natexlab{b}})\citenamefont{Varotsos, Sarlis, Skordas, Tanaka,
  and Lazaridou}}]{NAT06B}
\bibinfo{author}{\bibfnamefont{P.~A.} \bibnamefont{Varotsos}},
  \bibinfo{author}{\bibfnamefont{N.~V.} \bibnamefont{Sarlis}},
  \bibinfo{author}{\bibfnamefont{E.~S.} \bibnamefont{Skordas}},
  \bibinfo{author}{\bibfnamefont{H.~K.} \bibnamefont{Tanaka}},
  \bibnamefont{and} \bibinfo{author}{\bibfnamefont{M.~S.}
  \bibnamefont{Lazaridou}}, \bibinfo{journal}{Phys. Rev. E}
  \textbf{\bibinfo{volume}{74}}, \bibinfo{pages}{021123}
  (\bibinfo{year}{2006}{\natexlab{b}}).

\bibitem[{\citenamefont{Varotsos et~al.}(2009)\citenamefont{Varotsos, Sarlis,
  Skordas, and Lazaridou}}]{arxiv09a}
\bibinfo{author}{\bibfnamefont{P.~A.} \bibnamefont{Varotsos}},
  \bibinfo{author}{\bibfnamefont{N.~V.} \bibnamefont{Sarlis}},
  \bibinfo{author}{\bibfnamefont{E.~S.} \bibnamefont{Skordas}},
  \bibnamefont{and} \bibinfo{author}{\bibfnamefont{M.~S.}
  \bibnamefont{Lazaridou}} (\bibinfo{year}{2009}),
  \eprint{cond-mat/0707.3074v3}.


\bibitem[{\citenamefont{Varotsos and Alexopoulos}(1986)}]{newbook}
\bibinfo{author}{\bibfnamefont{P.}~\bibnamefont{Varotsos}},
  \emph{\bibinfo{title}{The Physics of Seismic Electric Signals}} (\bibinfo{publisher}{TerraPub},
  \bibinfo{address}{Tokyo}, \bibinfo{year}{2005}).



\end{thebibliography}

\end{document}